\documentclass[fleqn,usenatbib]{mnras}
\usepackage[T1]{fontenc}
\usepackage{ae,aecompl}

\usepackage{graphicx}
\usepackage{amsmath}	
\usepackage{amssymb}
\usepackage{siunitx}
\usepackage{subfigure}

\usepackage{color}
\usepackage{soul}
\usepackage[dvipsnames]{xcolor}

\definecolor{vero}{rgb}{0.5, 0.3, 0.7}
\definecolor{christi}{rgb}{0.0, 0.58, 0.71}
\definecolor{alexdu}{rgb}{0.00, 0.50, 0.00}
\definecolor{matt}{rgb}{0.8, 0.0, 0.0}

\newcommand{\lp}{ \left( }
\newcommand{\rp}{ \right) }
\newcommand{\lb}{ \left[ }
\newcommand{\rb}{ \right] }

\DeclareMathOperator\erf{erf}

\newcommand{\sifourfull}{Si~\textsc{iv} $\lambda\lambda$1393, 1402}
\newcommand{\cfourfull}{C~\textsc{iv} $\lambda\lambda$1548, 1550}

\newcommand{\sifour}{Si~\textsc{iv}}
\newcommand{\cfour}{C~\textsc{iv}}
\newcommand{\nfive}{N~\textsc{v}}

\newcommand{\ra}{R_\mathrm{A}} 
\newcommand{\rs}{R_{\ast}}
\newcommand{\ko}{\kappa_{0}}
\newcommand{\chii}{\chi_{\infty}}
\newcommand{\vinf}{v_{\infty}}
\newcommand{\vth}{v_{\mathrm{th}}}
\newcommand{\vlam}{v_{\lambda}}
\newcommand{\mdotb}{\dot{\mathrm{M}}_{\mathrm{B=0}}}

\newcommand{\xlos}{x_{\mathrm{LOS}}} 
\newcommand{\ylos}{y_{\mathrm{LOS}}} 
\newcommand{\zlos}{z_{\mathrm{LOS}}} 
\newcommand{\zb}{z_{\mathrm{B}}}

\newcommand{\kms}{km~s$^{-1}$}

\newcommand{\Alf}{Alfv\'en }

\defcitealias{Owocki2016}{O16}
\defcitealias{Hennicker2018}{H18}

\title[UVADM: UV Line Profiles of Massive Star Winds]{Ultraviolet Line Profiles of Slowly Rotating Massive Star Winds Using the ``Analytic Dynamical Magnetosphere'' Formalism}

\author[C. Erba et al.]
{C. Erba$^{1}$\thanks{E-mail: cerba@udel.edu}, 
A. David-Uraz$^{1,2,3}$,
V. Petit$^{1}$, 
L. Hennicker$^{4}$,
C. Fletcher$^{5}$,
A.W. Fullerton$^{6}$,
\newauthor
Y. Naz\'e$^{7}$,
J. Sundqvist$^{4}$, 
A. ud-Doula$^{8}$
\\
$^{1}$ Department of Physics and Astronomy, Bartol Research Institute, University of Delaware, Newark, DE 19716, USA \\
$^{2}$ Department of Physics and Astronomy, Howard University, Washington, DC 20059, USA \\ 
$^{3}$ Center for Research and Exploration in Space Science and Technology, and X-ray Astrophysics Laboratory, NASA/GSFC, Greenbelt, MD 20771, USA \\
$^{4}$ Institute of Astronomy, KU Leuven, Celestijnenlaan 200D, 3001 Leuven, Belgium \\
$^{5}$ Science and Technology Institute, Universities Space Research Association, Huntsville, AL 35805, USA \\
$^{6}$ Space Telescope Science Institute, Baltimore, MD 21218, USA \\
$^{7}$ Groupe d'Astrophysique des Hautes Energies, STAR, Universit\'e de Li\`ege, Quartier Agora (B5c, Institut d'Astrophysique et de G\'eophysique), All\'ee du 6 Ao\^ut 19c,\\ B-4000 Sart Tilman, Li\`ege, Belgium \\
$^{8}$ Department of Physics, Penn State Scranton, 120 Ridge View Drive, Dunmore, PA 18512, USA
}

\date{Accepted XXX. Received YYY; in original form ZZZ}

\pubyear{2021}

\begin{document}
\label{firstpage}
\pagerange{\pageref{firstpage}--\pageref{lastpage}}
\maketitle

\begin{abstract}
Recent large-scale spectropolarimetric surveys have established that a small but significant percentage of massive stars host stable, surface dipolar magnetic fields with strengths on the order of kG. These fields channel the dense, radiatively driven stellar wind into circumstellar magnetospheres, whose density and velocity structure can be probed using ultraviolet (UV) spectroscopy of wind-sensitive resonance lines. Coupled with appropriate magnetosphere models, UV spectroscopy provides a valuable way to investigate the wind-field interaction, and can yield quantitative estimates of the wind parameters of magnetic massive stars. We report a systematic investigation of the formation of UV resonance lines in slowly rotating magnetic massive stars with dynamical magnetospheres. We pair the Analytic Dynamical Magnetosphere (ADM) formalism with a simplified radiative transfer technique to produce synthetic UV line profiles. Using a grid of models, we examine the effect of magnetosphere size, the line strength parameter, and the cooling parameter on the structure and modulation of the line profile. We find that magnetic massive stars uniquely exhibit redshifted absorption at most viewing angles and magnetosphere sizes, and that significant changes to the shape and variation of the line profile with varying line strengths can be explained by examining the individual wind components described in the ADM formalism. Finally, we show that the cooling parameter has a negligible effect on the line profiles. 
\end{abstract}



\section{Introduction}

The ultraviolet (UV) spectra of hot, massive (O- and early B-type) stars include several wind-sensitive resonance lines that reveal the structure and kinematics of stellar winds. These spectra can therefore be coupled with wind models to quantify wind properties such as the mass-loss rate and terminal velocity.

To date, synthetic line profiles produced with spherically-symmetric wind models (e.g. \textsc{cmfgen}, \citealt{Hillier1998}) have been used to determine those properties for a large number of OB stars. However, certain physical phenomena such as rapid rotation \citep{Cranmer1995} and, of interest to this paper, the presence of large-scale surface magnetic fields, break down the assumed spherical symmetry in both wind density and flow velocity, thus affecting the shape of the UV line profiles. 

Recent spectropolarimetric surveys (MiMeS, BOB; \citealt{Morel2015,Wade2016,Grunhut2017}) have identified a distinct population of OB stars that host detectable surface magnetic fields. These fields channel the stellar wind into a circumstellar magnetosphere, confining it close to the stellar surface so that the stellar wind only escapes through open field lines. The closed magnetic loops co-rotate with the star, and when rotation is significant, it can provide centrifugal support to a part of the confined material. This forms a centrifugal magnetosphere \citep[CM;][]{Townsend2005a,Petit2013}. However, for low rotation rates, the trapped wind material continuously falls back to the stellar surface on a dynamical time scale, forming a dynamical magnetosphere \citep[DM;][]{Sundqvist2012,Petit2013}. \citet{Petit2013} classified magnetic OB stars into these two categories (CM vs. DM), and showed that the morphology of the H$\alpha$ line matches these characteristics. The majority of known magnetic O-type stars, and half of the known magnetic early-B stars, have DMs, which are the subject of this paper. 

For stars with DMs, the confinement of the stellar wind effectively reduces the rate at which mass is lost by the star as its wind escapes its gravity (when compared with a non-magnetic star of similar spectral type), with important evolutionary consequences \citep{Petit2017,Keszthelyi2019}. Even so, the material trapped within the magnetosphere still contributes significantly to the formation of the UV line profiles, but in a way that is distinct from a spherically symmetric outflow.  

Accordingly, the morphology of the UV resonance lines of magnetic O stars stands out compared to their non-magnetic counterparts. UV spectra of stars such as HD 108 \citep{Marcolino2012}, HD~191612 \citep{Marcolino2013}, CPD~-28{\textdegree}~2561 \citep{Naze2015}, and NGC~1624-2 \citep{DavidUraz2019,David-Uraz2021} show atypical profiles in many spectral lines (e.g. the \cfourfull \; and \sifourfull \; resonance lines), which can be qualitatively understood to result from the presence of a dipolar magnetic field \citep[][and above references]{Erba2017}. The spectra of these stars are further characterised by variability, which can be understood in the context of the Oblique Rotator Model \citep{Stibbs1950} as arising from the misalignment of the rotational and magnetic axes. This leads to rotational modulation. Therefore, for magnetic massive stars, synthetic line profiles produced with spherically symmetric wind models are often unable to successfully reproduce the shape of the observed line profile \citep{Marcolino2013,Erba2017,DavidUraz2019}, and furthermore yield unreliable estimates of wind properties.  

To address the challenges presented by the asymmetry of the magnetosphere, numerical magnetohydrodynamic simulations (MHD, e.g. \citealt{udDoula2002}) have been used to describe the density and velocity structure of the magnetically confined wind. When coupled with 3-dimensional radiative transfer techniques \citep[e.g.][]{Cranmer1996,Sundqvist2012}, UV line synthesis performed using the time-averaged output of these simulations has been successful in reproducing the character and shape of the line profile.

Such an analysis was performed by \citet{Marcolino2013}, who were able to qualitatively reproduce the variability observed in the \cfour \;and \sifour \;UV resonance lines of the magnetic O-type star HD~191612. A similar approach was adopted by \citet{Naze2015}, who used a 3D MHD simulation tailored to the magnetic O-type star HD~191612 (which was subsequently used by \citet{Naze2016}). This MHD model was coupled with the radiative transfer method from \citet{Sundqvist2012} and a \textsc{tlusty} photospheric profile \citep{Lanz2003} to produce synthetic UV line profiles of the O-type star CPD~-28{\textdegree}~2561. \citet{Naze2015} were able to reproduce the qualitative behavior of the \sifour~doublet with their synthetic line profiles. 

Despite these successes, MHD simulations have been unable to accurately reproduce the observed variability of the magnetic O-type star $\theta^1$~Ori~C, \citep[HD 37022;][]{Stahl1996,udDoula2008proc}. This method is also too computationally expensive for a large quantitative study, and is impractical for strong magnetic wind confinement due to the need for increasingly small Courant stepping times as the field strength is increased. It is therefore unsuitable to use for a systematic study of the many factors that affect UV line formation.

An alternative to MHD simulations for calculating the density and velocity structure of slowly rotating magnetospheres is provided by the Analytic Dynamical Magnetosphere (ADM) formalism \citep[][hereafter O16]{Owocki2016},
which has been shown to agree well with time-averaged 2D MHD simulations, and reproduces the observed H$\alpha$ variability of HD~191612 \citepalias{Owocki2016}.

An initial investigation of UV resonance line formation using the ADM formalism was presented in \citet[][hereafter H18]{Hennicker2018}. The authors used a 3D Finite Volume Method (3D-FVM) to discretize the equation of radiative transfer, and solved for the source function self-consistently using an accelerated $\Lambda$-iteration technique. Using four different combinations of the ADM formalism's description of the flow velocity and density within closed magnetic loops (see Section \ref{sec:ADM}), they produced several synthetic line profiles of a star with stellar, magnetic and wind parameters similar to those of HD~191612, which were shown to qualitatively compare well with those produced using an MHD simulation.
However, since the ADM formalism describes the time-averaged density structure at any given position in the magnetosphere as a superposition of upflow and downflow components, we note that the 3D-FVM method is unable to consider all of the components of the ADM model simultaneously. Furthermore, self-consistent 3D radiative transfer techniques accounting for supersonic velocity fields and line scattering are computationally demanding.
In this paper, we therefore use a more simplified radiative transfer scheme that was designed to consider all of the components of the ADM formalism simultaneously.

\citet{DavidUraz2019} used the ADM formalism, coupled with a modified Sobolev with Exact Integration (SEI) method for a singlet \citep{Hamann1981,Lamers1987} that used the optically thin source function (OTSF), to model the desaturation of the high-velocity edge of the absorption component of the high state\footnote{For magnetic massive stars, the term ``high state'' refers to variability in the H$\alpha$ spectral line; the peak of Halpha emission (and consequently the ``high state'') usually corresponds to the rotational phases for which the magnetic poles are the closest to our line of sight.} line profile of the magnetic O-type star NGC~1624-2. The authors made several simplifying assumptions in this model, including employing an infinite \Alf radius, and only using the upflow component of the wind (see Section \ref{sec:ADM}). \citet{DavidUraz2019} concluded that the resulting synthetic line profiles were able to reproduce the high velocity edge of the high state data from NGC~1624-2 with good agreement. 

Here, we present the first systematic parameter study of the formation of UV resonance lines in slowly rotating magnetic massive stars. We pair the ADM formalism with a simplified radiative transfer technique to produce synthetic UV line profiles that can be compared to observed spectra. In conjunction with our parameter study, we examine the effects of the individual components of the ADM formalism on the line profile. We also present the first complete application of the ADM formalism to a large magnetosphere.

In Section \ref{sec:methods}, we recapitulate the relevant ingredients of the ADM formalism, and discuss its implementation within a radiative transfer method. Section \ref{sec:results} discusses the morphology of several synthetic line profiles calculated for various typical model parameters, and discusses the link between the density and velocity structure predicted by the ADM formalism and the changes in the morphology of the synthetic line profiles. Finally, in Section \ref{sec:conclusions}, we summarize our findings and discuss future work.     


\begin{figure}
\centering
\includegraphics[width=0.45\textwidth]{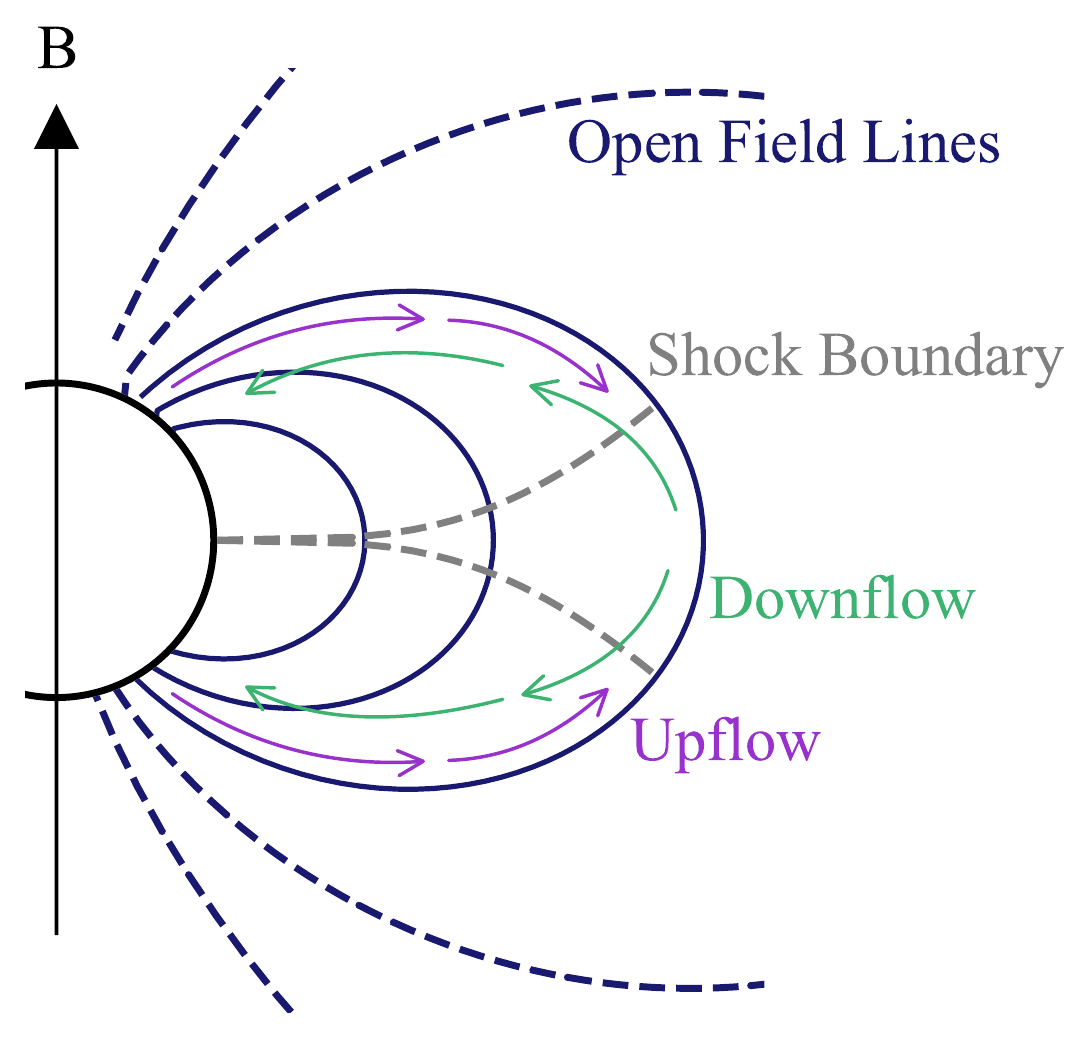}
\caption{Schematic of a star with a dipolar magnetic field (the magnetic axis corresponds to the black arrow) and its magnetosphere. The two components described by the ADM formalism that are important to UV radiative transfer are shown: the upflow (purple arrows) and the downflow (green arrows). The hot post-shock gas is located between the two shock boundaries (shown in grey), but does not contribute to the UV radiative transfer calculation. The open field region is considered to only contain upflowing material. A pole-on viewing angle ($\alpha = 0^{\circ}$) places the observer along the magnetic axis; an equator-on viewing angle ($\alpha = 90^{\circ}$) places the observer along the magnetic equator.}
\label{fig:toon}
\end{figure}

\section{Methods}
\label{sec:methods}

\subsection{The Analytic Dynamical Magnetosphere Formalism}
\label{sec:ADM}

The Analytic Dynamical Magnetosphere formalism \citepalias{Owocki2016} is a physically motivated, analytic description of the time-averaged mass flow within the closed magnetic loops of a centred dipolar magnetic field. It assumes that stellar rotation is not dynamically significant to the structure of the magnetosphere, which is adequate for most magnetic OB-type stars \citep{Petit2013}.

In the model, the mass flow within closed magnetic loops is divided into three components, illustrated in Figure \ref{fig:toon}:

\begin{itemize}
\item[(i)] The wind \textbf{upflow} consists of material that is radiatively driven from the stellar surface and is channeled along the field lines towards the magnetic equator; 
    
\item[(ii)] The \textbf{hot post-shock gas} is the result of the collision of upflowing material from each magnetic hemisphere. Because of the finite cooling length of the shock-heated plasma, for a given loop in each hemisphere, this region extends from a \textit{shock boundary} (dashed grey line in Figure \ref{fig:toon}) to the magnetic equator;
    
\item[(iii)] The wind \textbf{downflow} consists of the cooled post-shock material that flows back from the magnetic equator to the stellar surface under only the influence of stellar gravity.
\end{itemize}
    
The upflow and downflow wind are taken to co-exist at any point in space within the closed field lines. 
Realistically, within a single magnetic field tube, upflow and downflow wind material do not coexist, but because each field tube is independent and the dynamical time-scale of the wind is short (on the order of hours), a single snapshot in time averaged over 3D space yields a similar structure to that of a time-averaged MHD simulation \citep[see e.g.][]{Sundqvist2012, udDoula2013}. 

The magnitude of the upflow velocity is taken to be the canonical $\beta = 1$ velocity law ($ v/\vinf = 1 - \rs/r $ ), 
and is scaled in terms of the terminal velocity of the wind \citepalias[Equation 8]{Owocki2016}. The upflow speed is thus only a function of the radial coordinate; however, because the wind is ionized, the upflow direction follows that of the field lines. The upflow density is derived from the steady-state mass continuity equation \citepalias[Equation 9]{Owocki2016}.

In our implementation, the \Alf radius ($\ra$) marks the boundary between field lines that will remain closed (loops with a closure radius $<\ra$), and field loops that are ``open'' (loops with a closure radius $>\ra$). In the former case, the energy density of the magnetic field dominates the wind kinetic energy density, trapping the wind within the closed magnetic field lines. In the latter case, the wind kinetic energy density will dominate the field, forcing the magnetic field lines open and allowing the wind to escape (\citealt{udDoula2002}; see also the discussion on the ``deformed dipole'' topology in Section \ref{sec:adm_assumptions}). In the models discussed below in Section \ref{sec:results}, even though the field lines are considered open, we still maintain the dipolar geometry. The effect of this approximation on the UV line profiles is smaller in the case of strong magnetic confinement, and is discussed in Section \ref{sec:views}.

The location of the \Alf radius for a specific star can be approximated using the expression 
\begin{equation}
    \ra \sim \left[ \frac{B_{\mathrm{eq}}^2 \rs^2}{\mdotb \vinf} \right]^{1/4}
\end{equation}
where $B_{\mathrm{eq}}$ is the surface magnetic field at the equator, $\rs$ is the stellar radius, and $\mdotb$ and $\vinf$ are fiducial quantities representing the wind mass-loss rate and terminal speed the star {\em would have} in the absence of a magnetic field \citep{udDoula2002}. The quantity $\mdotb$ is also referred to as the \textit{wind-feeding rate}, in order to distinguish it from e.g. the integrated surface mass flux (which depends on the inclination of the magnetic field to the stellar surface) or the ``real'' mass-loss rate (the amount of material that actually escapes through open field lines; \citealt{DavidUraz2019}).   

Inside the closed magnetic loops, the upflow region terminates at the shock boundary. This location is determined by solving the transcendental equation given in Equation B16 of \citet{udDoula2014}. Note that the shock-heated gas, with temperature generally in excess of $10^6$ K, is essentially {\em transparent} to UV line scattering, due to the increased ionization of the associated atomic species; therefore, we ignore this region for the UV radiative transfer. 

The location of the shock boundary depends on the cooling parameter $\chii$, which is related to the radiative cooling length \citep{udDoula2014}:
\begin{equation}
\chii = 0.034 \frac{V_8^4 R_{12}}{\dot{M}_{-6}},
\label{eq:chii}
\end{equation}
where $R_{12} \equiv \rs/10^{12}$ cm, $V_8 \equiv \vinf/10^8 \; \textrm{cm} \; \textrm{s}^{-1}$, and $\dot{M}_{-6} \equiv \mdotb/10^{-6} \; \textrm{M}_{\odot} \; \textrm{yr}^{-1}$. If the cooling parameter is large, the hot post-shock gas covers a wider area around the magnetic equator, resulting in a decreased contribution to the line profile from the upflow wind (see Sec. \ref{sec:views} for a more in-depth discussion of this effect).

In the ADM model, the wind downflow results from the shocked gas that cools and slows while traversing the shock region, and then
falls back to the stellar surface starting at the magnetic equator. The downflow density is given by \citetalias[Equation 23]{Owocki2016}, which we modify here to express the ratio of the downflow density $\rho_{\mathrm{c}}$ in terms of the same fiducial density as the upflow $\rho_{\mathrm{w_{\ast}}}$:

\begin{equation}
\frac{\rho_{\mathrm{c}}}{\rho_{\mathrm{w_{\ast}}}} = \frac{\rho_{\mathrm{c}}}{\rho_{\mathrm{c_{\ast}}}} \frac{\vinf}{v_{\mathrm{e}}}, 
\end{equation}
where $v_{\mathrm{e}}$ is the escape speed\footnote{The value of $(\vinf/v_{\mathrm{e}})$ is empirically either 1.3 (on the cool side of the bi-stability jump) or 2.6 (on the hot side of the bi-stability jump; \citealt{Vink2001}). In our case, the latter is the appropriate choice \citep{Lamers1995}, but for simplicity we round to $(\vinf/v_{\mathrm{e}}) = 3$. Our modeling shows that choosing a value of $(\vinf/v_{\mathrm{e}}) = 1.3$ would significantly increase the amount of red absorption present in the line profile at all viewing angles.}. The downflow velocity can also be recast in terms of the terminal velocity from \citetalias[Equation 22]{Owocki2016}. We note there is no downflow density component in the magnetic loops outside of the \Alf radius.

Additionally, we employ a smoothing length $\delta = 0.1$ (as illustrated in \citetalias[Figure 4]{Owocki2016}) to spatially smooth out the downflow region, thus avoiding a singularity at the magnetic equator. 


\subsection{Radiative Transfer}
\label{sec:rt}

Our radiative transfer calculation uses a 3D Cartesian grid in the frame of reference of the observer, where the line-of-sight axis $\zlos$ is in the direction of the observer.

The magnetosphere reference frame is oriented toward the observer by a right-handed rotation about an arbitrarily defined $\xlos$ axis (perpendicular to the $\zlos$ axis) by viewing angle $\alpha$, where $\cos \alpha = \hat{z}_{\mathrm{B}} \cdot \hat{z}_{\textrm{LOS}}$, and the magnetic moment vector lies along the $\zb$ axis. The angle $\alpha$ therefore describes the angle between the line-of-sight to the observer and the north magnetic pole.

For a star with obliquity $\beta$ (between the rotation axis and the magnetic axis), inclination $i$ (between the rotation axis and the line-of-sight axis), and rotational phase $\phi$, the viewing angle is given by \citep{Stibbs1950}:
\begin{equation}
    \cos \alpha = \sin \beta \cos \phi \sin i + \cos \beta \cos i .
\end{equation}
For a dipole viewed ``pole-on,'' $\alpha = 0 \si{\degree}$, and for a dipole viewed ``equator-on,'' $\alpha = 90 \si{\degree}$. 
   
In the case considered here, where the rotation is slow enough that it does not dynamically impact the structure of the magnetosphere, the phase variation of the line profile can be completely described by a single magnetospheric structure viewed from different values of $\alpha$ \citep{udDoula2008,Sundqvist2012}. The short-term variability caused by dynamic motions in the magnetosphere was shown to be small in H$\alpha$ \citep{udDoula2013}, but has not been investigated in the UV. In our models, we only consider viewing angles between $0-90 \si{\degree}$, because of the north-south symmetry of a centred dipole about the magnetic equator. 

We use a uniform grid in $(\xlos, \ylos)$ space (spanning the plane perpendicular to the observer's line-of-sight) with range $[-10 \rs, 10 \rs]$, sampling $N_x$ = $N_y$ = 401 for a total of 160,801 spatial rays.
As the spatial evaluation of the ADM model values is relatively fast, we calculate the ADM values directly along a given ray, as opposed to interpolating rays at a certain viewing angle though a pre-computed magnetosphere.
Rays that intersect the stellar surface at coordinate $z_{\ast}$ start with a continuum specific intensity $I_{\ast}$. 
Rays that do not intersect the stellar surface are initiated in the model at $\zlos = -10 \rs$ with $I(\zlos = -10 \rs) = 0$. We note that we do not consider limb darkening in this model, therefore $I_{\ast}$ is uniform over the stellar disk. Additionally, the models discussed here present only the wind component -- no photospheric line profile is included in the computation of the line profiles discussed below. Given the assumed slow surface rotation of these stars, a photospheric profile would only span the central part of the line, and so would not have a significant impact on the velocity range of the full line profile.

The wavelength coordinate, defined in velocity space $\lp \vlam \rp$, is scaled to the terminal speed. Indeed, within the ADM formalism all velocities are expressed in terms of the terminal speed, thus a comparison with data requires the scaled $\vlam$ to be converted back to velocity units (e.g. km\,s$^{-1}$) using the terminal speed the star would have if no magnetic field was present \citep{udDoula2002}. The terminal velocity of the wind is therefore an indirect free parameter of the synthetic UV spectra when performing a direct comparison with observations. We use a grid in 
velocity space of $N_{\lambda}$ = 49 points spread uniformly in Doppler velocity space about line centre from $v/\vinf=\lb-1.2,1.2\rb$ in order to sample the entire width of the line profile.

We reiterate that along a given ray, upflow and downflow wind material is considered simultaneously within the closed loops (the optical depths are added at corresponding grid points). Within the post-shock region, the hot gas and the downflow wind technically coexist; however, since the hot gas is essentially transparent to UV line scattering, only the downflow wind is considered in this regime. This is in contrast to the method presented by \citetalias{Hennicker2018}, who employ four different spatial combinations of the upflow and downflow wind (see discussion below). In the open field regions, only upflow is considered. 

We solve for the specific intensity $I(x,y,\vlam,\tau=0)$ as follows. For a constant source function $S$ between two spatial coordinates (corresponding to two adjacent grid points) $\lb z_{\textrm{LOS, n}}, z_{\textrm{LOS, n+1}} \rb$ along a ray, the specific intensity is given by
\begin{equation}
I(\tau_{\mathrm{n+1}}) = I(\tau_{\mathrm{n}}) e^{\tau_{\mathrm{n+1}}-\;\tau_{\mathrm{n}}} + S\left[ 1-e^{\tau_{\mathrm{n+1}}-\;\tau_{\mathrm{n}}} \right]
\label{eq:formsol}
\end{equation}
where the optical depth $\tau_{\mathrm{n}} > \tau_{\mathrm{n+1}}$ (thus n increases toward the observer). The first term on the right-hand side of the equation determines the contribution from absorption, while the second term determines the contribution from emission. We assume single resonant line scattering processes with isotropic redistribution, and so set the source function equal to the mean intensity \citep{Owocki1996}. The limit for the optically thin regime ($\tau \lesssim 1$) is assumed, such that 

\begin{equation}
\label{eq:sf}
\frac{S}{I_{\ast}} = \frac{\bar{J}}{I_{\ast}} = \frac{ 1 - \sqrt{ 1 - \lp \frac{\rs}{r} \rp^2 } }{2},
\end{equation}
where $r$ is the radial distance from the centre of the star to the point at which the source function is being calculated. Note that here the source function is independent of wavelength ($\vlam$), as we assume a constant photospheric continuum across the wavelength range of the line profile. The applicability of the optically thin source function to UV line profile synthesis is discussed further in Section \ref{sec:otsf_proof}.

Following the same notation as in Equation \ref{eq:formsol}, the change in optical depth between two successive grid points is given by 
\begin{equation}
\tau_{\mathrm{n+1}} - \tau_{n} =  - \int_{z_{\textrm{LOS, n}}}^{z_{\textrm{LOS, n+1}}} k(\vlam, z) \rho(z) dz .
\label{eq:taudef}
\end{equation}

We assume the density $\rho(z)$ is constant over each integration step 
$\lb z_{\textrm{LOS, n}}, z_{\textrm{LOS, n+1}} \rb$. The velocity is set to vary linearly over the same interval. We choose this piecewise linear velocity approach to the integration due to the small width of the profile function compared to the range of velocities along the ray. This approach ensures the variation of the optical depth within the resonance zone(s) is well sampled, without having to enforce an artificially large spatial resolution that could lead to computationally expensive integration times.

The line opacity $k(v_{\lambda},z) = \ko \Phi \lp \vlam, z \rp$ can be expressed using the dimensionless line strength parameter \citep{Hamann1980,Sundqvist2014}:
\begin{equation}
    \ko = \frac{\mdotb \; q}{\rs \vinf^2} \frac{\pi e^2 / m_{\mathrm{e}} c}{4 \pi m_{\mathrm{H}}} \frac{a_{\mathrm{i}}}{1 + 4Y_{\mathrm{He}}} f_{\mathrm{lu}} \lambda_0,
\label{eq:kappa_0}
\end{equation}
where $m_{\mathrm{H}}$ is the mass of Hydrogen, $c$ is the speed of light, and $e$ and $m_{\mathrm{e}}$ are the charge and mass of an electron respectively.  
The value of $\ko$ depends on a specific elemental transition through the ion fraction $q$, the abundance of the element with respect to hydrogen $a_{\mathrm{i}}$, the helium number abundance $Y_{\mathrm{He}}$, the rest wavelength $\lambda_0$, and the oscillator strength $f_{\mathrm{lu}}$. 
Note that the ADM formalism does not incorporate a method to determine the relevant ion fraction in the magnetosphere; to compute $\ko$ for a specific star and spectral line, we would estimate the ion fraction from 1D non-local thermodynamic equilibrium (NLTE) codes e.g. \textsc{cmfgen}.

The local line profile is approximated by a Gaussian function that reflects the underlying thermal Doppler broadening produced by a 1-D Maxwellian distribution of the velocity of the atoms along the line of sight:
\begin{equation}
\label{eq:prof_func}
\Phi\lp \vlam, z \rp = \frac{1}{\vth} \frac{1}{\sqrt{\pi}} \exp\left[-\left(\frac{\vlam-v(z)}{\vth}\right)^2\right]
\end{equation}
Here, $v(z)$ is the line-of-sight velocity of the wind at the specific location sampled, and $\vth$ is the thermal velocity, for which we choose the value $\vth = 0.01 \vinf$\footnote{For typical O-type star parameters, $\vth$ is on the order of 10 km\,s$^{-1}$. For more specific comparisons (e.g. to observed spectra), this parameter can be adjusted to a more precise value (see Section \ref{sec:microturb}).}. For cases where the slope of $v(z)$ approaches zero, we take a second order Taylor Expansion of Equation \ref{eq:prof_func} about the zero-point of the slope. 

Equation \ref{eq:taudef} for the optical depth is therefore
\begin{equation}
\tau_{\mathrm{n+1}} - \tau_{\mathrm{n}} = - \frac{\ko \rho}{\vth} \int_{z_{\mathrm{n}}}^{z_{\mathrm{n+1}}} \frac{1}{\sqrt{\pi}} \exp\left[-\left(\frac{\vlam-v(z)}{\vth}\right)^2\right] dz .
\end{equation}
Here, $\rho$ is taken to be the density at the mid-point between subsequent grid points along ray ($\rho = \lb \rho(z_{\mathrm{n}}) + \rho(z_{\mathrm{n+1}}) \rb/2$), and $\ko$ is assumed to be constant over the spatial step. Replacing v(z) with a linear interpolation between two grid points, 
\begin{equation}
v(z) = \lp \frac{ v(z_{\mathrm{n+1}}) - v(z_{\mathrm{n}}) }{ z_{\mathrm{n+1}} - z_{\mathrm{n}} } \rp \lp z - z_{\mathrm{n}} \rp + v(z_{\mathrm{n}}),    
\end{equation}
we obtain the analytical solution:
\begin{equation}
\label{eq:tau_singleslab}
\tau_{\mathrm{n+1}} - \tau_{\mathrm{n}} = \frac{\ko \rho}{2} \left( \frac{z_{\mathrm{n+1}} - z_{\mathrm{n}}}{v(z_{\mathrm{n+1}})-v(z_{\mathrm{n}})} \right) (\erf(u(z_{\mathrm{n}})-\erf(u(z_{\mathrm{n+1}})), 
\end{equation}
where $u \lp z \rp = (\vlam - v(z) )/\vth$. We finally solve for $I(\tau = 0)$ by sweeping Equation \ref{eq:formsol} along the ray, with the change in optical depth given by Equation \ref{eq:tau_singleslab}. 

\begin{figure}
\centering
\includegraphics[width=0.47\textwidth]{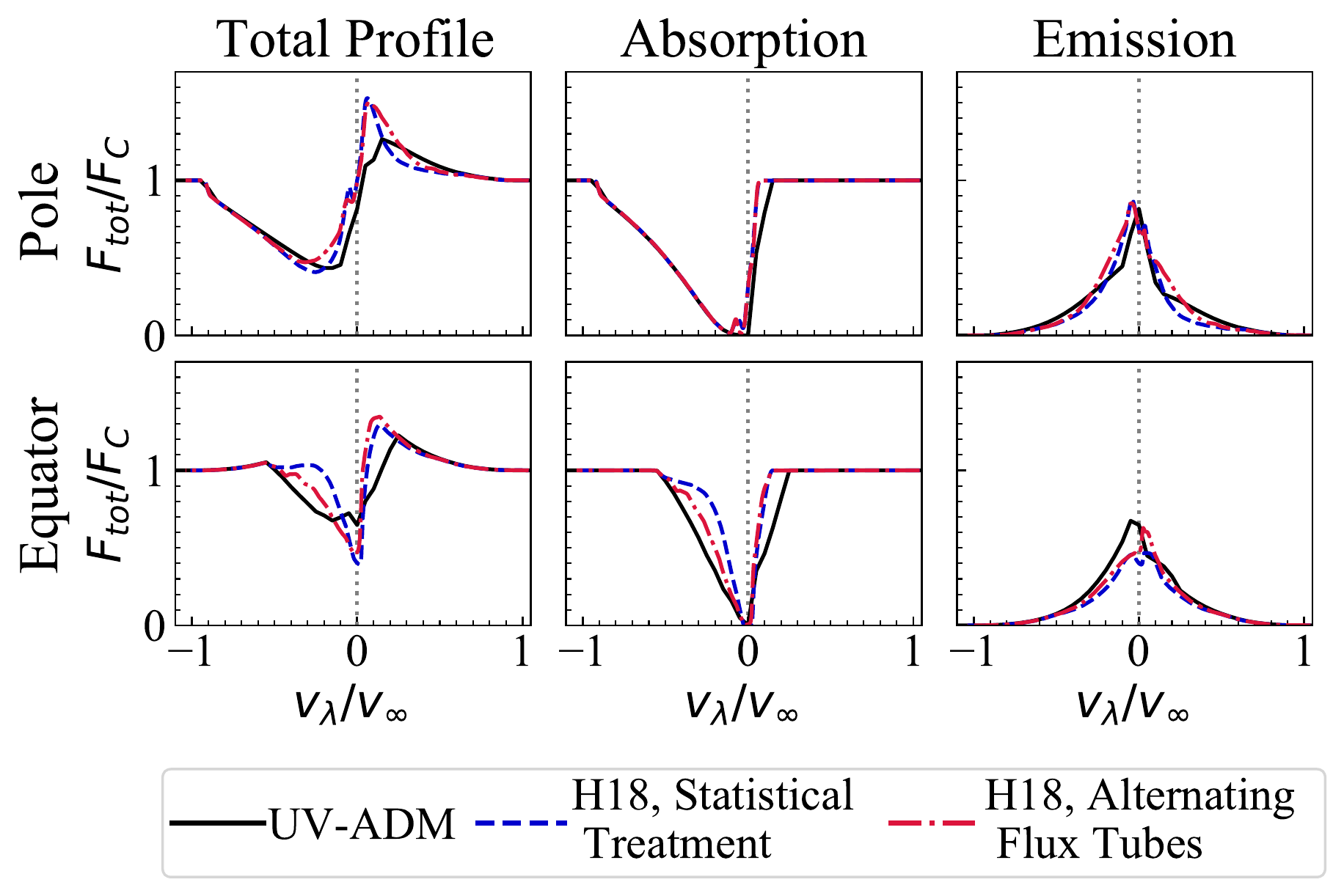}
\caption{Synthetic UV line profiles calculated with the UV-ADM code (black solid lines) of a star with model parameters ($\ra = 2.7~\rs, \ko = 1.0, \chii = 0.01$), similar to those used for the ADM-type profiles in Figures 11 and B.1 of \citetalias{Hennicker2018}. We also show the ``statistical treatment'' (\citetalias{Hennicker2018}, Model ii; blue dashed lines) and the ``alternating flux tubes'' (\citetalias{Hennicker2018}, Model iii; red dot-dashed lines) line profiles for comparison. The left column shows the total line profiles, the middle column shows the absorption profiles, and the right column shows the emission profiles for a pole-on (upper row) and equator-on (lower row) view of the magnetosphere.}
\label{fig:h2018comp}
\end{figure}

Figure \ref{fig:h2018comp} shows synthetic line profiles calculated with the UV-ADM code (black solid lines), with similar parameters to those used in the ADM-type models of \citetalias{Hennicker2018}. We also show for comparison the ``statistical treatment'' (model ii; blue dashed curves) and the ``alternating flux tubes'' (model iii ; red dot-dashed curves) line profiles from \citetalias{Hennicker2018} (see their Figures 11 and B.1). In that work, the authors show their models compare qualitatively well with synthetic line profiles calculated from MHD simulations. 

The UV-ADM model in Figure \ref{fig:h2018comp} uses a small value of $\chii = 0.01$, since \citetalias{Hennicker2018} do not include shock retreat within their calculations. We also scale our synthetic line profiles by a factor of 1.5 in velocity\footnote{In MHD simulations, the observed polar flow velocity is higher than the terminal velocity the model would have in the absence of the magnetic field. This was explained by a faster-than-radial expansion of the wind above the poles, leading to a desaturation of the blue edge of the line, and therefore allowing for further radiative driving.} \citep{Owocki2004}, to match their method, although we do not apply this correction factor in the rest of this paper. Within the confined region, we consider the upflow and downflow material simultaneously, whereas \citetalias{Hennicker2018} use four different methods for combining the upflow and downflow material because 3D-FVM only allows for one value of the density to be considered at a given location. As Figure \ref{fig:h2018comp} demonstrates, although we use a more simplistic radiative transfer scheme, we obtain similar line profiles overall to those shown by \citetalias{Hennicker2018}.

We finally note that the ``downflow only'' model from \citetalias{Hennicker2018} considers a case where there is downflow material within the closed loops {\em and} upflow material in the open loops. In Section \ref{sec:results}, we illustrate the separate contribution of the upflow and downflow wind components to the total line profile. In our work, the line profiles labelled as ``downflow'' consider only the contribution of the downflow material confined within closed loops (that is, upflow material within open loops is not included). Therefore, our downflow-only profiles should not be compared directly to the downflow-only profiles from \citetalias{Hennicker2018}, because they do not illustrate the same case.

\subsection{Is the Optically Thin Source Function Sufficient?}
\label{sec:otsf_proof}

As discussed in Section \ref{sec:rt}, the line opacity (Equation \ref{eq:kappa_0}) is highly dependent on the wind mass-loss rate and on the atomic parameters of the line. In typical O-type stars with $\dot{M} \approx 10^{-6}$ M$_{\odot}$ yr$^{-1}$, this can lead to relatively large line strength parameters (e.g. $\ko = 30$ for \cfour~in $\zeta$ Pup; \citealt{Hamann1980}). Because the optical depth depends on the line opacity (see Equation \ref{eq:taudef}), a large line strength parameter calls into question the suitability of the optically thin limit ($\tau \lesssim 1$) applied in our simplified calculation of the source function (Equation \ref{eq:sf}).

\begin{figure}
\centering
\includegraphics[width=0.45\textwidth]{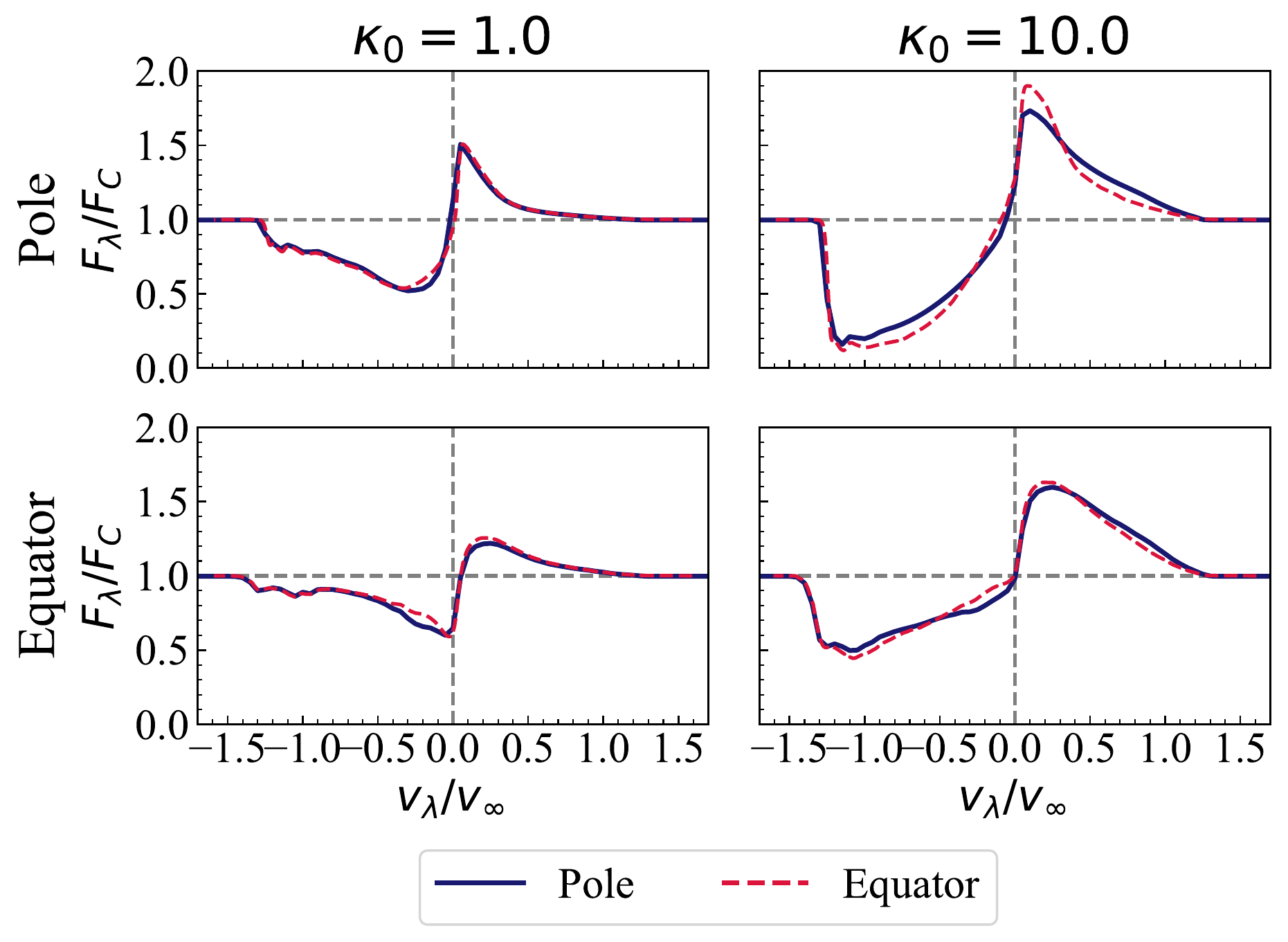}
\caption{A snapshot of the 3D MHD model of the magnetosphere of $\theta^1$~Ori~C is coupled with the radiative transfer technique from this paper (which uses the optically thin source function, blue solid lines), and the short characteristics method from \citet[][red dashed lines]{Hennicker2020}. Both sets of models were calculated for a magnetic pole-on (upper row) and equator-on (lower row) view, using line strength parameters of $\ko = 1.0$ (left column) and $\ko = 10.0$ (right column), and $\theta^1$~Ori~C characteristics ($\ra = 2.3~\rs, \; \vinf = 3200$ \kms). The two methods agree well.}
\label{fig:otsf_compare}
\end{figure}

We investigate this question by producing two sets of line profiles using a 3D MHD model of the magnetosphere of $\theta^1$~Ori~C \citep{udDoula2013}. The resulting line profiles are shown in Figure \ref{fig:otsf_compare}. The first set is generated by coupling the MHD magnetosphere with the radiative transfer technique from this paper (using the optically thin source function), while the second set applies the short characteristics method from \citet{Hennicker2020}, which solves for the source function self-consistently. Both sets of line profiles are computed for parameters appropriate to $\theta^1$~Ori~C ($\ra = 2.25~\rs, \; \vinf = 3200$ \kms), line strength parameters of $\ko = 1.0$ (left column) and  $\ko = 10.0$ (right column), for a magnetic pole-on (upper row) and equator-on (lower row) view. We choose an MHD simulation (instead of a model generated using the ADM formalism) for this comparison, in order to circumvent differences in the line profiles that may arise from the upflow and downflow densities in the ADM model that cannot be treated simultaneously by the short characteristics method. 

The agreement between the two methods is quite good, with only minimal discrepancies appearing near line centre. At the typical signal-to-noise ratios expected for hot star UV spectroscopy, these differences would likely be indistinguishable. The largest line strength parameter presented in this paper is $\ko=1.0$,\footnote{The choice of a line strength parameter of $\ko=1.0$ for modeling UV resonance line profiles occurs several times in the literature, and is therefore a useful choice for comparison. \citet{Marcolino2013} used $\ko = 1.0$ to model ``moderately strong'' lines in HD~191612 (O6.5f?pe-O8fp; \citealt{Howarth2007}). In their investigation of UV resonance line formation in CPD~-28{\textdegree}~2561 (O6.5f?p; \citealt{Walborn2010}), \citet{Naze2015} chose $\ko = 1.0$ to model a ``generic singlet line'' as a proxy for overlapping doublets (e.g. \nfive~or \cfour). 
We stress that the value of $\ko$ is highly dependent on the individual line in a specific star, so there is no ``one size fits all'' approach to assessing which line strength parameters correspond with particular lines. The line opacity, and consequently the applicability of the optically thin source function, needs to be uniquely evaluated for each line/star combination in any direct comparison of synthetic and observed UV spectra.} therefore this comparison demonstrates that the optically thin source function is sufficient at least to this limit. Additionally, as Figure \ref{fig:otsf_compare} shows, even for an order-of-magnitude larger line strength of $\ko = 10.0$, 
the mean relative difference of the line profiles calculated with the OTSF and the short characteristics method remains small. We note that the future development of a model grid with line strength parameters larger than this limit may require a reevaluation of the applicability of this approximation.


\section{Variations in the Line Profile}
\label{sec:results}

\begin{figure*}
\centering
\subfigure[$\ra = 2.7~\rs$]{
\includegraphics[scale=0.58]{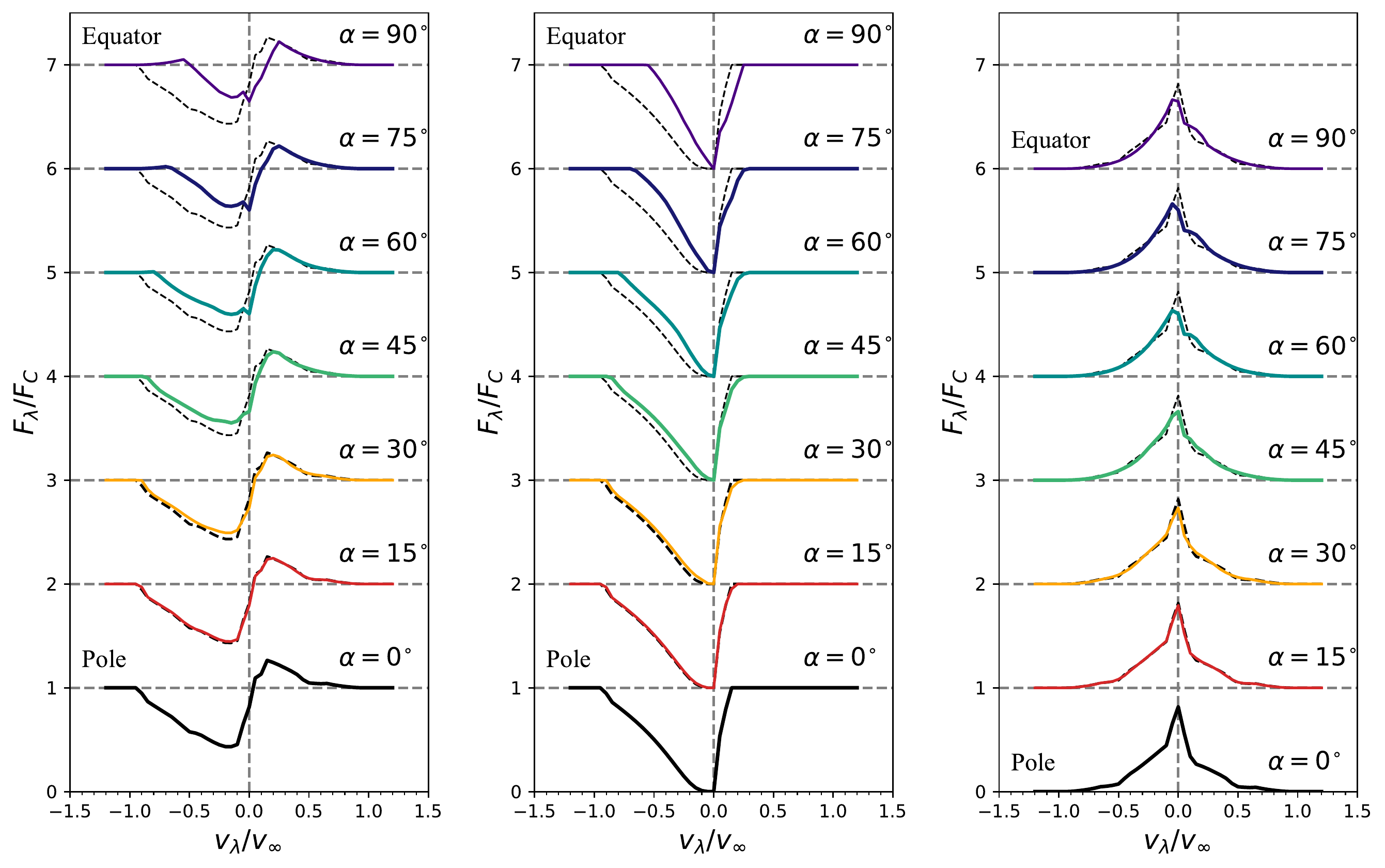}
\label{fig:phases_2p7}
}
\subfigure[$\ra = 10.0~\rs$]{
\includegraphics[scale=0.58]{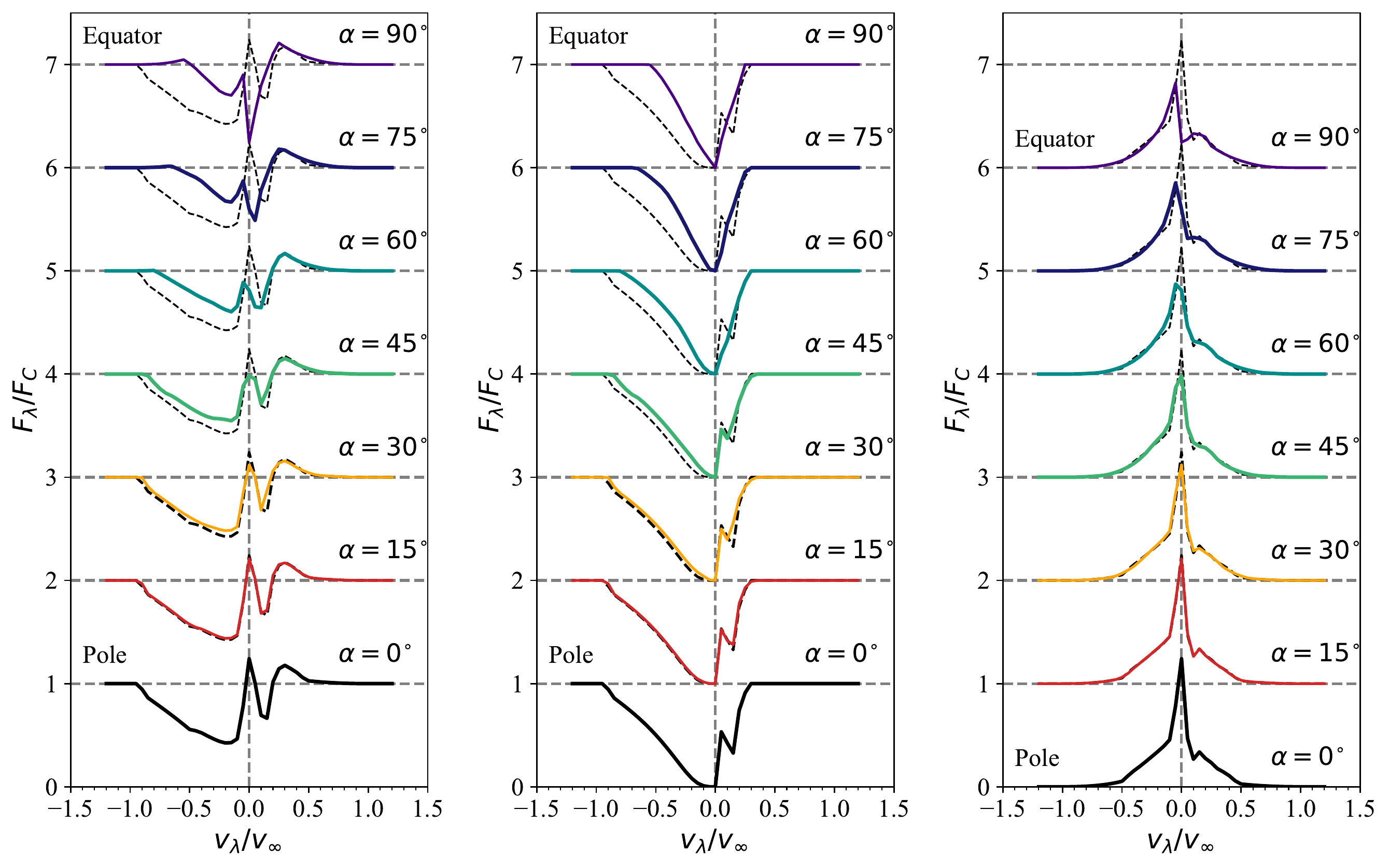}
\label{fig:phases_10}
}
\caption{Synthetic UV line profiles calculated at different viewing angles, with \Alf radius $\ra=2.7~\rs$ (top) and $\ra=10.0~\rs$ (bottom), cooling parameter $\chii=1.0$, and line strength $\ko = 1.0$. Each subfigure contains three panels showing the full line profile (left), as well as the individual absorption (middle) and emission (right) components. The black dashed line at each viewing angle replicates the pole-on profile for comparison, demonstrating the modulation observed as the star rotates.}
\label{fig:phases}
\end{figure*}

The following parameter study is built on values that are chosen to roughly correspond to particular well-known magnetic massive stars. For simplicity, the present study only addresses singlet lines, in order to highlight the variation of an individual line profile with changing physical parameters, and to avoid the complexities associated with doublets. As many of the important wind lines present in the current UV spectra available for magnetic O-type stars are doublets (see Figure 3 of \citealt{DavidUraz2019}), we postpone a direct comparison between our synthetic spectra and observations to a forthcoming paper. 

In Section \ref{sec:views}, we explore the effects of the viewing angle on the line profile for a magnetosphere with two different \Alf radii. We choose (a) $\ra=2.7~\rs$, similar to the O-type star HD~191612, and used in \citet{Marcolino2013} for their UV synthetic spectra calculated from MHD simulations, and (b) $\ra=10.0~\rs$, similar to NGC~1624-2, the most strongly magnetic O-type star observed to date \citep{Wade2012}. These models have a line strength parameter of $\ko=1.0$, corresponding to a moderately strong line (e.g. \cfour, for the typical O star mass-loss rate that we consider in this study), and a moderate cooling parameter of $\chii = 1.0$ (similar to HD~191612). 

In Section \ref{sec:line_strength}, we address the impact of the line strength parameter on the line profile by revisiting the models considered in Sec. \ref{sec:views} for a weak line strength parameter of $\ko=0.1$ (corresponding to e.g. \sifour). Similarly to the analysis presented in Sec. \ref{sec:views}, we extend the previous discussions of the effect of $\ko$ on the line profile by examining the impact of the individual upflow and downflow wind components on the (separated) absorption and emission profiles. 

Section \ref{sec:cool_param} considers the effect of the cooling parameter on the line profile for each of the models presented in Sections \ref{sec:views} and \ref{sec:line_strength}, at extremum values for a low ($\chii = 0.01$) and high ($\chii = 100$) cooling parameter. We also address here the effect of the smoothing length parameter on the downflow wind material for the models presented in Section \ref{sec:views}.

Finally, Section \ref{sec:microturb} considers the impact of increasing the thermal velocity term within the profile function (in order to approximate the effect of a turbulent velocity dispersion; see Equation \ref{eq:prof_func}) on the line profiles from Section \ref{sec:views}.


\subsection{Viewing Angle and \Alf Radius}
\label{sec:views}

\begin{figure}
\centering
\includegraphics[width=0.45\textwidth]{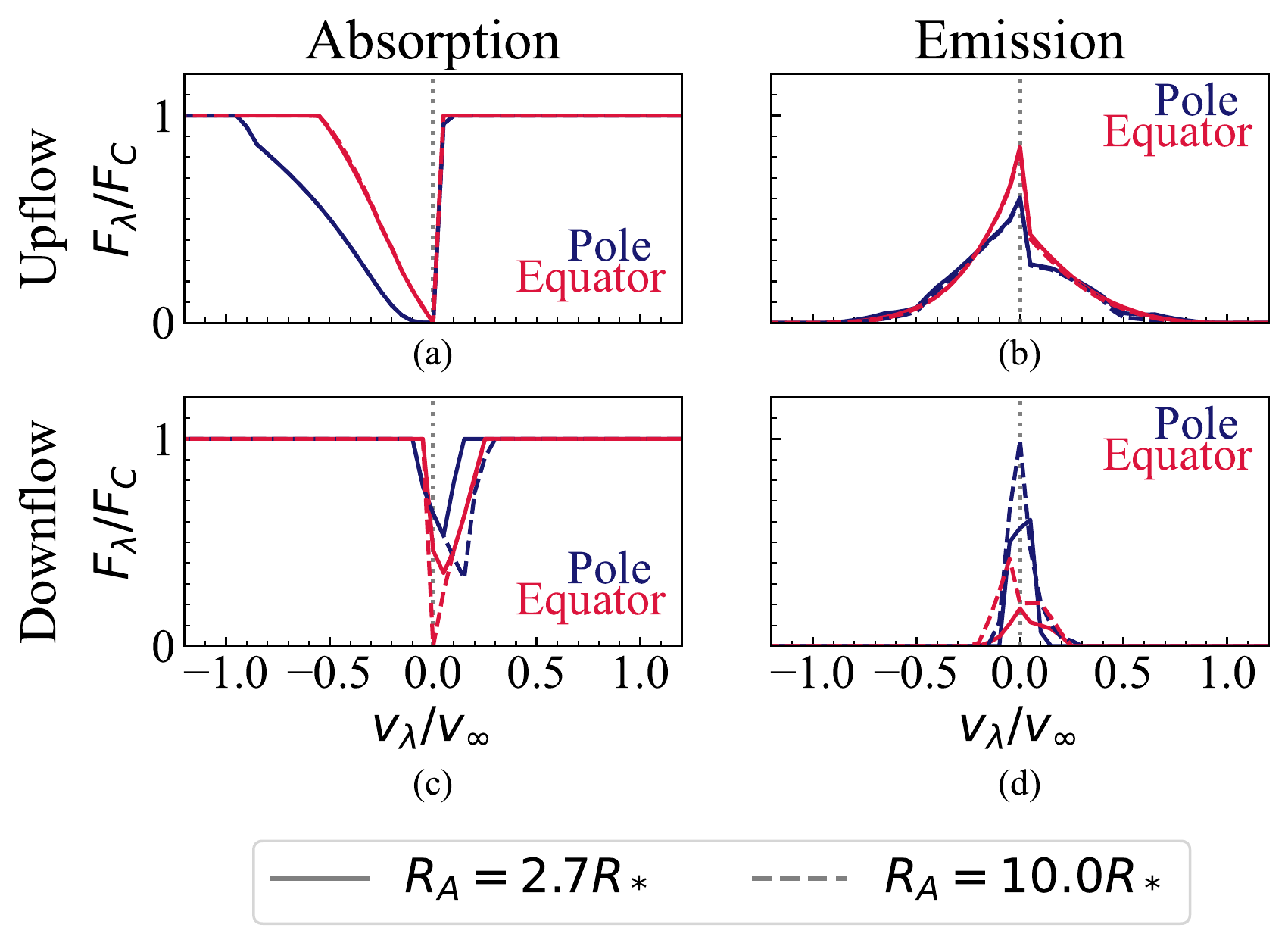}
\caption{Synthetic UV absorption and emission profiles calculated at pole-on and equator-on viewing angles, with \Alf radii of $\ra=2.7~\rs$ and $\ra=10.0~\rs$, cooling parameter $\chii=1.0$, and line strength $\ko = 1.0$. The upflow (top row) and downflow (bottom row) components show absorption (left) and emission (right) profiles originating exclusively from the upflow or downflow wind. The line profile features formed by the upflow and downflow are distinct, emphasizing their separate contributions to the shape of the total line profile.} 
\label{fig:updown}
\end{figure}

Figure \ref{fig:phases} illustrates the variation with viewing angle of the line profiles, for two magnetospheres with different \Alf radii (a. $\ra=2.7~\rs$ and b. $\ra=10.0~\rs$). Each panel shows an evenly-spaced progression in $\alpha$ from a pole-on view ($\alpha = 0^{\circ}$) to an equator-on view ($\alpha = 90^{\circ}$). For comparison, the pole-on line profile is also displayed by a dashed line in the panels for viewing angles $\alpha > 0^{\circ}$. The left-hand column in each subfigure shows the full line profile, while the middle and right columns show the individual absorption and emission components, respectively (see Equation \ref{eq:formsol}). 
\begin{figure*}
\centering
\subfigure[$\ra = 2.7~\rs$]{
\includegraphics[width=0.45\textwidth]{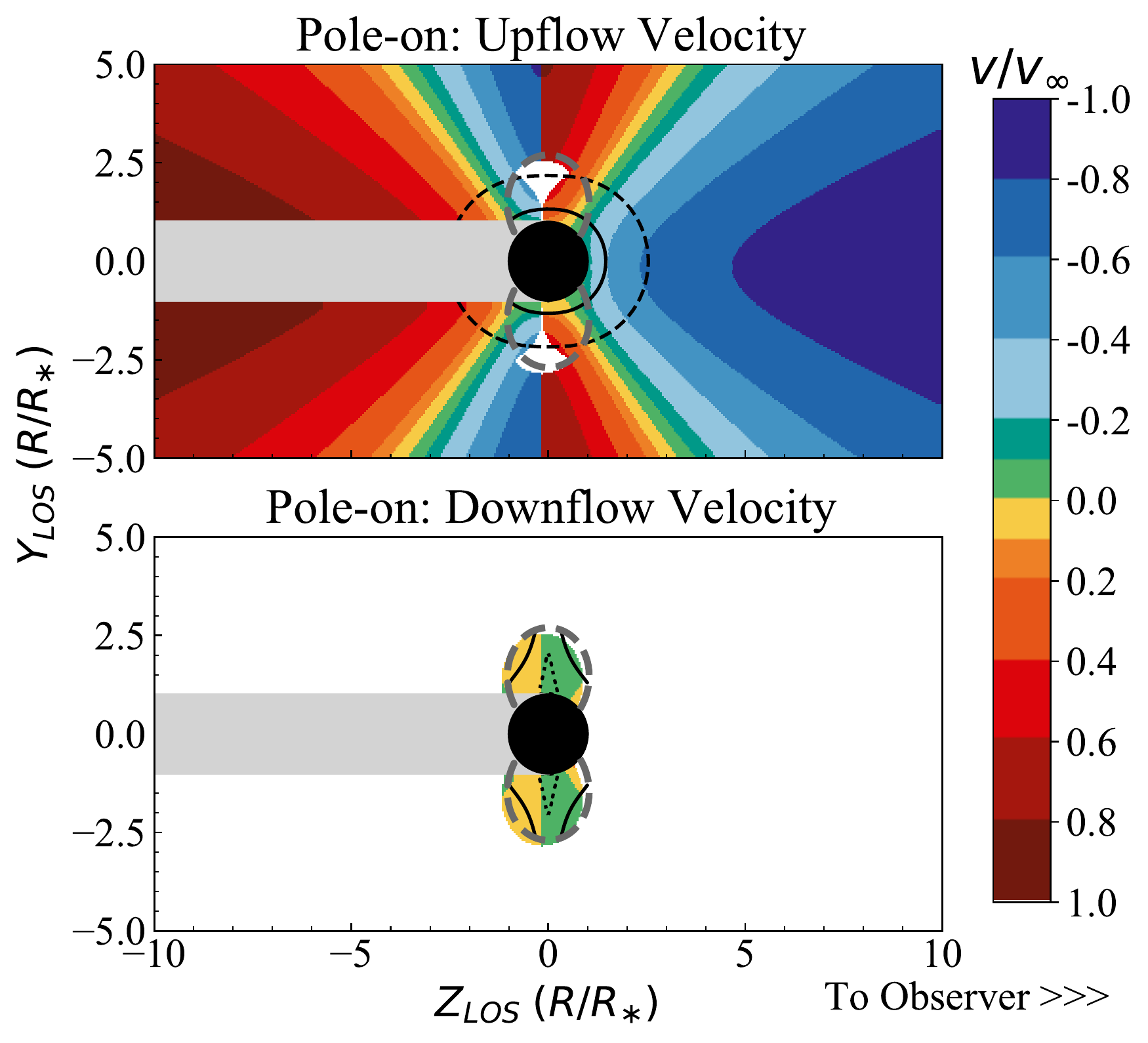}
\label{fig:vpole_2p7}
}
\hspace{0.5cm}
\subfigure[$\ra = 10.0~\rs$]{
\includegraphics[width=0.45\textwidth]{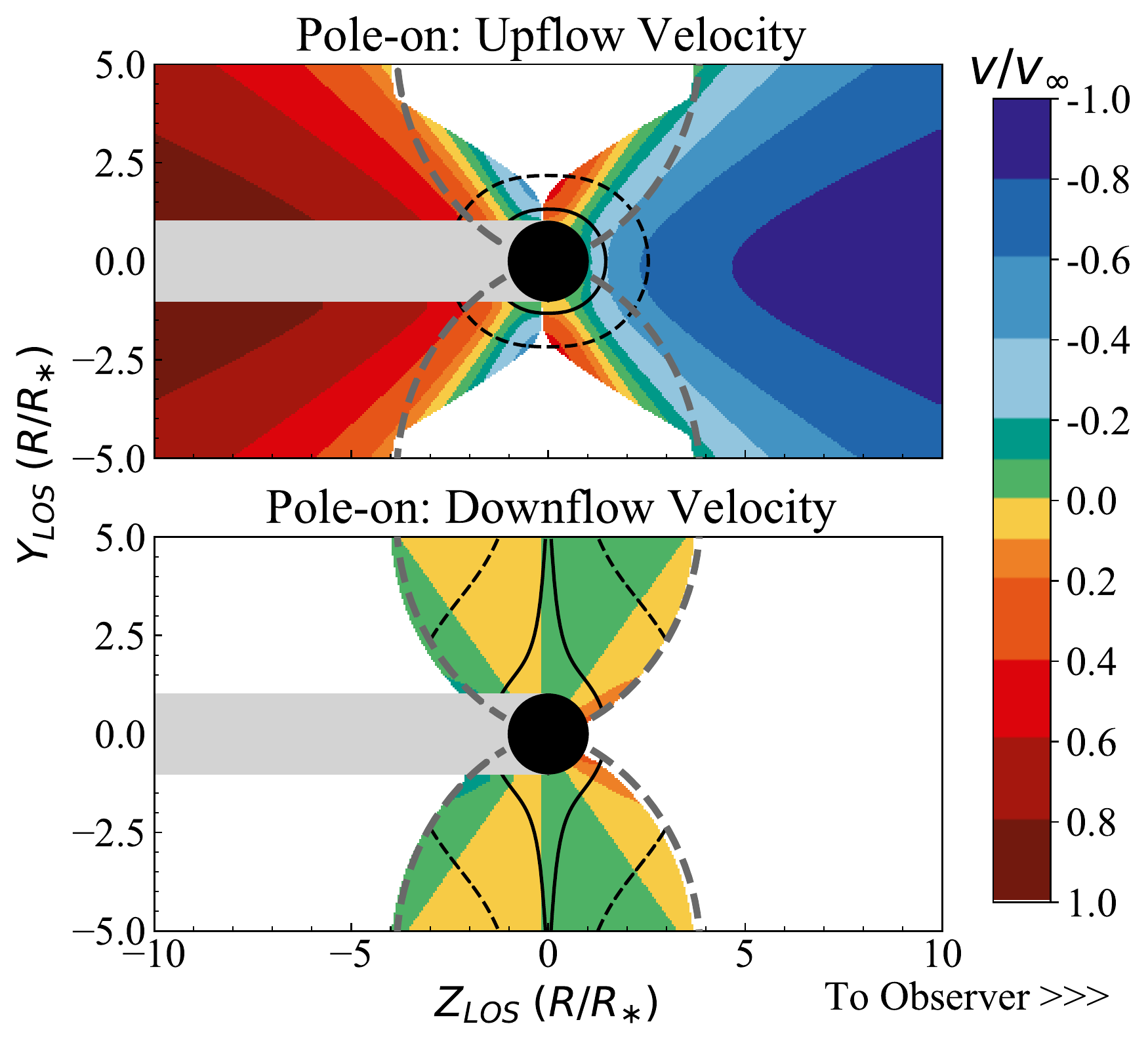}
\label{fig:vpole_10}
} \\
\subfigure[$\ra = 2.7~\rs$]{
\includegraphics[width=0.45\textwidth]{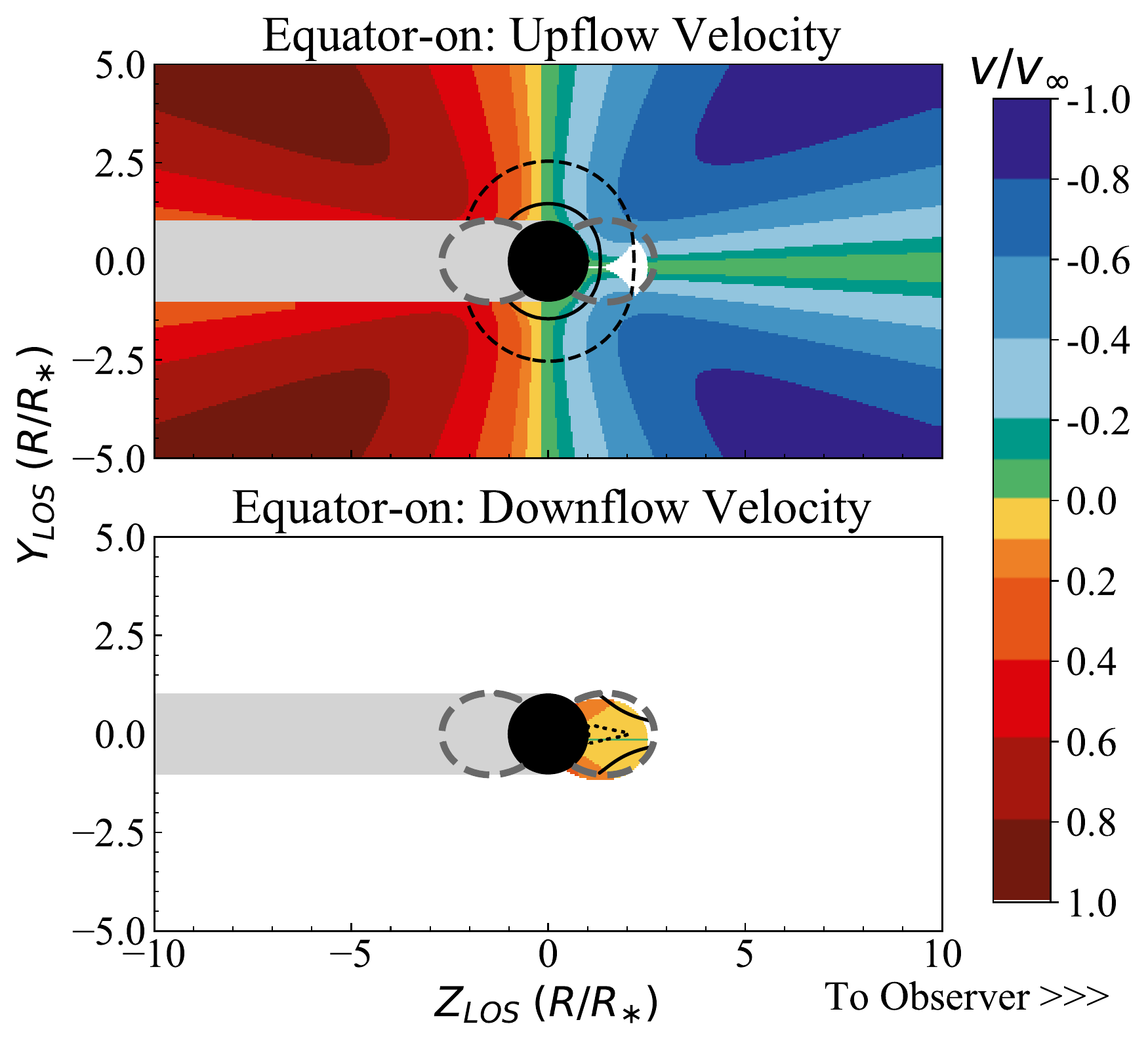}
\label{fig:veq_2p7}
}
\hspace{8pt}%
\subfigure[$\ra = 10.0~\rs$]{
\includegraphics[width=0.45\textwidth]{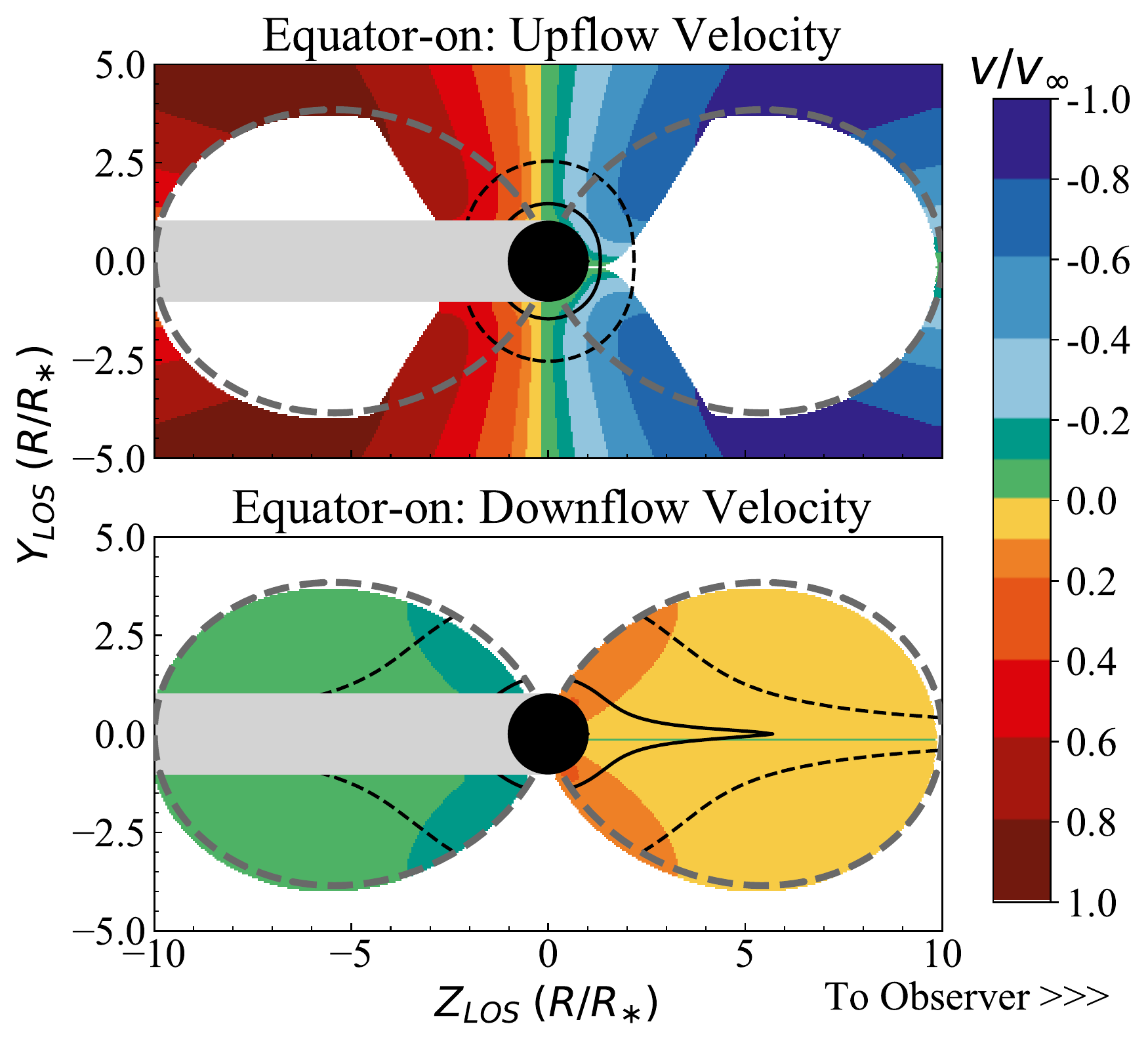}
\label{fig:veq_10}
}
\caption{Line-of-sight velocity maps for \Alf radius $\ra=2.7~\rs$ and $\ra=10.0~\rs$, and cooling parameter $\chii=1.0$, illustrated by a slice in the $\xlos = 0$ plane. The grey dashed lines represent the location of the last closed magnetic loop, and the observer is located to the right in each image. Density contours are outlined in black, with solid lines at $\rho/\rho_{\mathrm{w_{\ast}}} = 1.0$, dashed lines at $\rho/\rho_{\mathrm{w_{\ast}}} = 0.1$, and dotted lines at $\rho/\rho_{\mathrm{w_{\ast}}} = 5.0$. Upflow wind is truncated at the shock boundary, which does not affect the downflow material. The white shading inside the closed magnetic loops indicates the location of the hot post-shock gas.} 
\label{fig:vplot}
\end{figure*}

The shape and variation of the line profile at different viewing angles can be understood as follows: the absorption component is due to the intervening column of material between the stellar disk and the observer, scattering photospheric light out of the line of sight. This absorption column is composed of {\em both} upflow and downflow material. To understand the separate roles of these ADM components in shaping the total line profile, Figure \ref{fig:updown} (left column) shows the absorption profiles due to the upflow (top) and downflow (bottom) components, for the two \Alf radii (solid and dashed lines), at pole-on (dark blue lines) and equator-on (red lines) viewing angles. Additionally, Figure \ref{fig:vplot} shows a contour map of the line-of-sight velocity in the plane containing both the line-of-sight and magnetic axes, for the same configurations as Figure \ref{fig:updown}, illustrating the geometry of the magnetically channeled upflow and downflow wind. Each panel of Figure \ref{fig:vplot} also shows two density contours to illustrate regions of high and low density around the star.         

Below, we discuss in turn the behavior of the blue side and the red side of the line.

\textbf{(i) Blue side of the line}

For a pole-on ($\alpha = 0^{\circ}$) view, most of the absorption due to the upflow occurs blueward of line centre. This is illustrated in Figures \ref{fig:vpole_2p7} and \ref{fig:vpole_10} by the green (blue) colors, representing low (high) blueshifted velocities in the absorption column in front of the star. The downflow material also contributes to the absorption profile (see Figure \ref{fig:updown}, panel c, blue lines). Field loops that close below $\sim1.8~\rs$ have a minimal contribution from the downflow wind to the absorption at low velocity blueward of line centre (when compared to that of the upflow material). 

Figure \ref{fig:updown} (panel a) shows that while the blueshifted absorption extends to the terminal velocity for the pole-on view, the absorption is less extended for the equator-on view. In both cases, the absorption is saturated at line centre. The former effect is reflected in the variation of the total line profile, whereas the latter is offset by the emission. 

Finally, Figure \ref{fig:updown} also illustrates that the contribution of the upflow wind to the absorption profile is not dependent on \Alf radius (dashed vs. solid curves), due to the approximation that the unconfined upflow still follows the dipole field lines (see Sec. \ref{sec:adm_assumptions}).

\textbf{(ii) Red side of the line}

A part of the upflow wind at the pole-on viewing angle is directed away from the observer. This material is located close to the limb of the stellar disk, and is caused by magnetic loops that close below 1.8~$\rs$. Figure \ref{fig:updown} (panel a, dark blue lines) shows this contribution is insignificant to the shape of the total line profile.  

More importantly, {\em if} the magnetosphere has $\ra \gtrsim 1.8~\rs$, the downflow material in loops that close at or above 1.8~$\rs$ will contribute to the absorption redward of line centre. Therefore, the contribution of the downflow to the red absorption is stronger for a magnetosphere with a larger \Alf radius (Figure \ref{fig:updown}, panel c, blue solid and dashed lines). 

The variation of the redshifted absorption with viewing angle is only due to the downflow component.
Figures \ref{fig:veq_2p7} and \ref{fig:veq_10} show that there is more low-velocity downflow material (yellow color) in the absorption column for the equator-on view, leading to a stronger overall red absorption. 
However, for a magnetosphere with a large \Alf radius, the shallower absorption redward of line centre for the pole-on view extends to larger redshifted velocities. This is because the absorption column includes high-velocity downflow near the stellar surface (Figure \ref{fig:vpole_10}, orange color) for the pole-on view, whereas the absorption column consists of mostly low-velocity downflow at the top of the loops (Figure \ref{fig:veq_10}, yellow color) for the equator-on view.

Redshifted absorption is discernible at all viewing angles in the total line profile for magnetospheres with large \Alf radii, but is hidden at low viewing angles for magnetospheres with small \Alf radii. 

In a star with a spherically symmetric stellar wind, the flow is directed radially away from the stellar surface; consequently, the absorption column will never contain plasma moving away from the observer. Redshifted absorption in the line profile is a distinct signature of the presence of a magnetic field, and has been observed in the \cfour~and \sifour~doublets of NGC~1624-2 at low state (corresponding to an approximately equator-on view for this star; \citealt{DavidUraz2019}), as well as in HD 54879 (viewing angle unknown; \citealt{Shenar2017,DavidUraz2019}). 

Furthermore, large periodic variation of the blue side of the line profile is also strongly suggestive of a large-scale magnetic field, although such variability could also be ascribed to other wind structures, such as co-rotating interaction regions \citep{Cranmer1996,DavidUraz2017}.


The emission component of the line profile is formed by photospheric radiation scattered into the line of sight of the observer. A spherically symmetric stellar wind therefore results in a nearly symmetric emission profile with respect to line centre, with a small amount of redshifted emission missing from the line profile due to the occultation of the rear hemisphere by the stellar disk. In contrast, the presence of a magnetic field introduces asymmetries in the emission profile that cannot exclusively be explained by occultation. 

The right-hand panels of Figure \ref{fig:phases_2p7} and Figure \ref{fig:phases_10} show the emission line profiles for $\ra = 2.7~\rs$ and $\ra = 10.0~\rs$, respectively. In contrast to the broad and smooth emission profile resulting from a spherically symmetric wind,\footnote{Except for the small discontinuity introduced by the occultation of the wind by the stellar disk.} the emission profiles shown here are broad at higher velocities with a narrow peak near line centre. This central peak is from the downflow wind's contribution to the line profile -- because the downflow velocity never exceeds the defined $v_{e}/\vinf$ for the model, the downflow is the cause of the emission peak at low velocities.

Our investigation revealed that the symmetry of the emission part of the line profile about line centre is a complex function of the magnetospheric geometry, in contrast to the simple explanation above for a spherically symmetric wind. The complex line-of-sight velocity structure results in the possibility of crossing multiple resonance zones along a given ray. Even if the line-of-sight velocity structure in the forward hemisphere of the magnetosphere is the mirror image of the negative of the velocity structure in the rear hemisphere, the relative observed intensities at the same $|\vlam|$ will be different depending on the local value of the source function at the last resonance zone encountered, which is dependent on the radial distance from the star. This can be seen, for example, in Figure \ref{fig:vpole_2p7}, where a ray at $\ylos = 2.5$ will cross through $v/\vinf = 0.3$ (orange color) twice, once in the backward hemisphere and once in the forward hemisphere of the magnetosphere. In this example, the resonance zone in the backward hemisphere is crossed first but is further from the star, therefore the value of the source function -- and of the intensity -- will be smaller when crossing the first resonance zone than when crossing the second.

This said, Figure \ref{fig:updown} (panels b and d) shows that the variation of the emission part of the line profile with viewing angle is small, compared to that of the absorption. Thus in the total line profiles, the low-velocity emission peak contributes to the atypical shape of the line profile (especially for magnetospheres with large \Alf radii), but the variation with viewing angle is mostly driven by the absorption. 


\subsubsection{Effect of ADM Assumptions on the Line Profile}
\label{sec:adm_assumptions}

\begin{figure}
\centering
\includegraphics[width=0.45\textwidth]{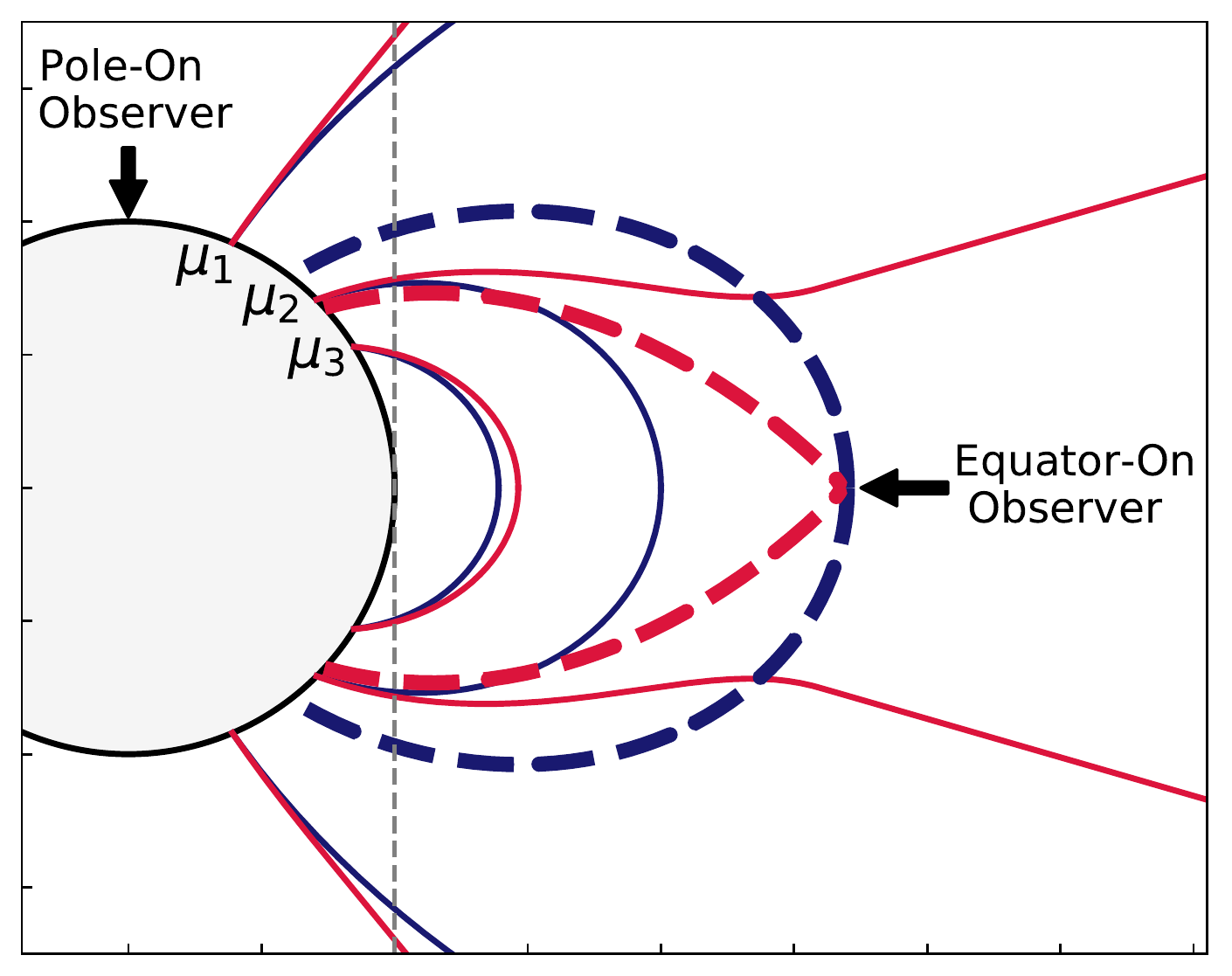}
\caption{Schematic of a star with a purely dipolar magnetic geometry (blue curve) and with a deformed dipole topology (red curve), where the field lines become radial near the \Alf radius. We indicate the three footpoints of the separate magnetic loops discussed in Section \ref{sec:adm_assumptions} using $\mu_1$, $\mu_2$, and $\mu_3$. The last closed magnetic loops are shown with thick dashed curves for both cases. The \Alf radius of the schematic is set to $ \ra = 2.7~\rs$; the discrepancy between the different field topologies is less distinct at larger \Alf radii.}
\label{fig:adm_fieldextract}
\end{figure}

The ADM formalism makes the assumption that the magnetic field topology is dipolar everywhere in the magnetosphere. Outside of closed magnetic loops, the wind (upflow) is assumed to follow the direction of the dipolar field lines. In reality, the wind kinetic energy density overcomes the magnetic energy density, such that loops that would have an apex above the \Alf radius are opened, and do not necessarily follow a dipolar topology. Also, closed loops with an apex near to (but still below) the \Alf radius are also deformed with respect to a purely dipolar topology
\citep[see e.g. Figure~1 of][]{udDoula2013}. In this ``deformed dipole'' topology, the wind direction in the open-field region would become radial in the vicinity of the \Alf radius. Such a configuration would have little effect for a pole-on view of the magnetosphere, where the open field lines are already nearly radial; however, for an equator-on view, the change in field geometry could have a significant impact on the flow direction.

This issue was highlighted by \citetalias{Hennicker2018} as the primary source of the discrepancies between synthetic line profiles produced using the ADM formalism and those produced using MHD simulations. They noted that the deformed dipole configuration would more closely resemble the MHD results for an equator-on view, and concluded that the ADM formalism as-is does not accurately model the open field region. 

We further note that some magnetic massive stars have surface field topologies that are not dipolar (e.g. $\tau$ Sco [B0.2V], \citealt{Donati2006b}; HD 37776 [B2V], \citealt{Thompson1985}). In such cases, the ADM formalism's assumption of a dipolar geometry would be inappropriate. However, the ADM model can be adapted to a field of arbitrary shape \citep{Fletcher2018}, and so can be extended to non-dipolar topologies.

While such non-dipolar topologies and the results of MHD simulations are outside the scope of this paper, we provide here a qualitative assessment of the impact of open field lines for a surface dipolar field. To mimic this process, we calculate the magnetic field in the magnetosphere assuming a potential field with a dipolar surface boundary condition and an outer boundary condition set such that the field becomes radial at the \Alf radius
(the so-called source surface). We calculate the resulting deformed dipole magnetic field and field lines following the method presented in \citet{Jardine1999} and \citet{Donati2006b}.

Figure \ref{fig:adm_fieldextract} shows these two topologies (the pure dipole field in blue, the deformed dipole model in red) for a star with $\ra = 2.7~\rs$. The last closed magnetic loops are shown with thick dashed curves for both cases. Note that although they have the same closure radius, they do not have the same footpoint $\mu = \cos \theta$ (where $\theta$ is the colatitude) at the surface of the star. For the deformed dipole topology, the closed loops cover a smaller volume in the magnetosphere and cover a smaller area at the surface of the star, so there is less material in the confined regions. The ADM formalism thus overestimates the amount of red absorption due to the downflow at all phases, particularly at low velocities near line centre. Most of the downflow wind is at low velocities with respect to the upflow, so the change in flow direction due to the deformation of the field lines should only have a small impact on the line profile.  

Figure \ref{fig:adm_fieldextract} also shows the distortion of field lines with the same footpoints $\mu$ (thin curves) and hence the same local outward surface mass flux. 
The field lines with footpoint $\mu_1$ are located within the pole-on absorption column. The deformed dipole loops only diverge from the pure diople loop far from the stellar surface, at which point the upflow density does not significantly contribute to the optical depth. Therefore, the blue absorption for the pole-on view will remain largely unchanged.  

The field loops with footpoint $\mu_2$ are within the equator-on absorption column. This loop is closed for the pure dipole and open for the deformed dipole. The change of field line direction in the equator-on absorption column is therefore significant.
As a reminder, the velocity extent of the blue absorption for an equator-on view is smaller than for a pole-on view. 
For a deformed dipole, the field lines near the magnetic equator but above the \Alf radius will contribute blue absorption at higher velocities than for the pure dipole case, leading to a more extended blue wing in the total line profile. We would thus expect the modulation of the blue absorption with stellar rotation to be lessened; however, the density is low in this region.

Additionally, closed field loops in the deformed dipole geometry with footpoints near $\mu_2$ have a more peaked shape near the loop apex than the same field loops in the pure dipole geometry (see e.g. Figure \ref{fig:adm_fieldextract}, thick dashed lines). In theory, this could desaturate the absorption at line centre in the equator-on view: wind material following the more peaked curvature of the deformed dipole loops has more line-of-sight velocity than in the pure dipole case (in which the wind material has a near-zero line-of-sight velocity near loop apex). However, in the equator-on view, much of the upflow wind that would be in this region is also within the extent of a shock boundary, and is therefore not contributing to the UV line profiles (see Figures \ref{fig:vplot} and \ref{fig:duckplot}, and the discussion in Section \ref{sec:cool_param}). Furthermore, for larger magnetospheres, the downflow wind density is low in this region (see Figure \ref{fig:veq_10}, dashed black lines), and so does not significantly affect the absorption profile. Therefore, the effect on the absorption profile of this change in shape of the field lines in the equator-on view is negligible in all cases except in small magnetospheres with very small ($\chii \sim 0.01$) cooling parameters.

The field loops at footpoint $\mu_3$ illustrate the negligible distortion of the deformed dipole case compared to the pure dipole case for loops far inward of the closure radius, hence the effects of a deformed dipole will be less pronounced. As the \Alf radius becomes larger, the field lines will be very similar to that of a pure dipole in regions where the density is high enough to be significant to the opacity. Thus, for $\ra \sim 10 \rs$ or greater (see density contours in Figure \ref{fig:vplot}), the dipole field approximation is adequate.

\begin{figure}
\centering
\includegraphics[width=0.45\textwidth]{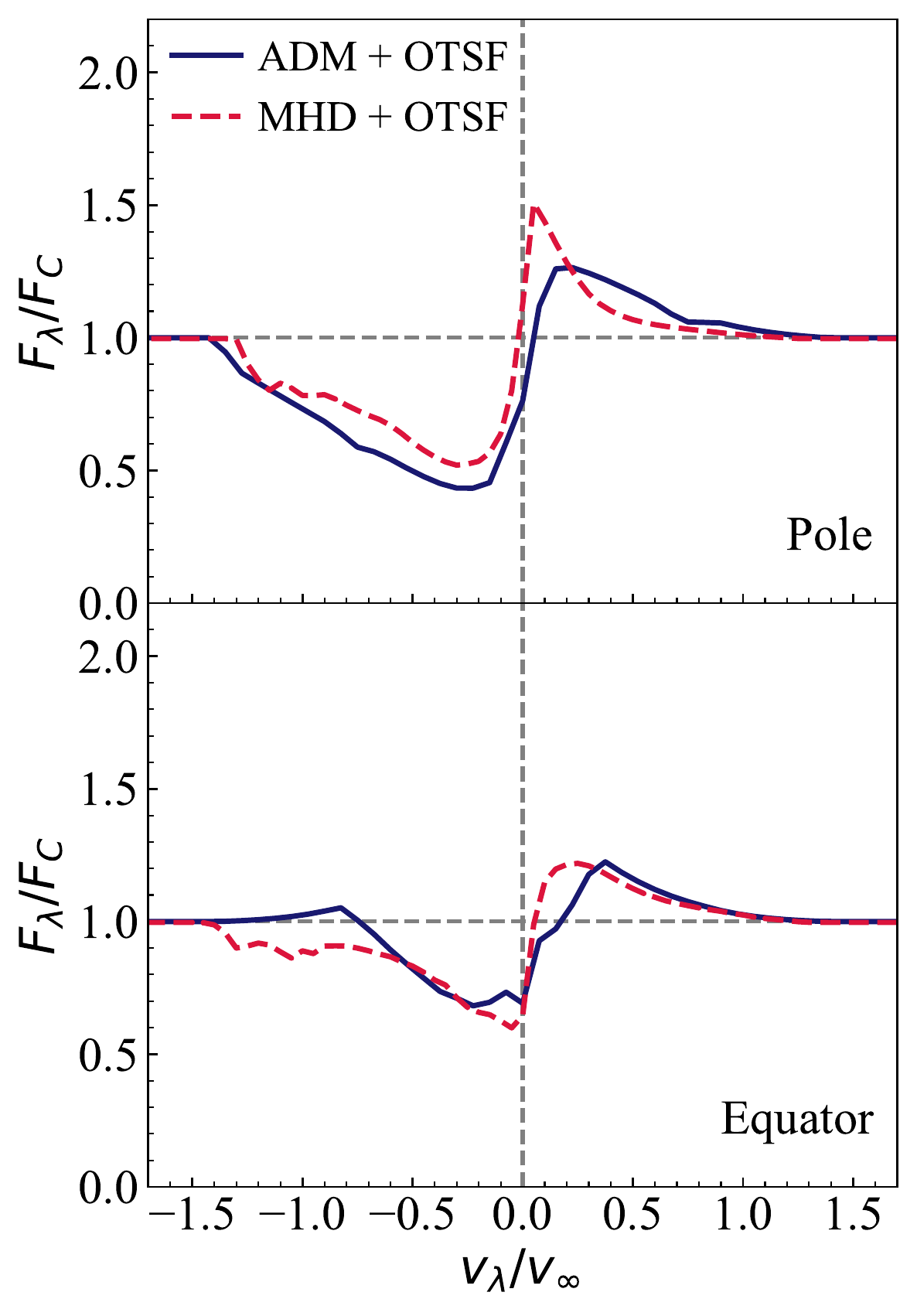}
\caption{A synthetic UV line profile calculated using the ADM formalism (blue solid line), compared with the 3D MHD snapshot of the magnetosphere of $\theta^1$~Ori~C from Figure \ref{fig:otsf_compare}. The MHD model has been coupled with the radiative transfer technique from this paper that uses the optically thin source function (red dashed lines). Both sets of models were calculated for a magnetic pole-on and equator-on view, using a line strength parameter of $\ko = 1.0$, and $\theta^1$~Ori~C characteristics ($\ra = 2.3~\rs, \; \vinf = 3200$ \kms). The qualitative shape of the line profiles agrees well for small $\ra$, which is where the greatest difference between line profiles produced using the MHD and ADM methods would be expected.}
\label{fig:otsf_mhdadm}
\end{figure}

As a proof-of-concept example, we show in Figure \ref{fig:otsf_mhdadm} the pole-on and equator-on line profiles from Figure \ref{fig:otsf_compare} generated using the 3D MHD model of the magnetosphere of $\theta^1$~Ori~C \citep{udDoula2013}, coupled with the radiative transfer method from this paper (using the optically thin source function). We compare this to a set of line profiles calculated using the ADM formalism, with model characteristics similar to $\theta^1$~Ori~C ($\ra = 2.3~\rs, \; \vinf = 3200$ \kms), coupled with the same radiative transfer technique. As in Figure \ref{fig:h2018comp}, we have scaled the synthetic line profiles calculated using the ADM magnetosphere by a factor of 1.5 in velocity in order to provide a proportionate comparison to the line profiles calculated from the MHD simulation.

In general, the two methods have qualitatively similar morphologies: the overall shape of the total line profile is similar, and the rotational modulation between a pole-on and an equator-on view of the magnetosphere is observed in both sets of line profiles. There are some discrepancies: in the pole-on view, the synthetic line profiles produced using the ADM formalism slightly underestimate the emission at redshifted velocities near line centre. This is consistent with the results reported by \citetalias{Hennicker2018}, who performed a similar comparison using ADM and MHD magnetospheres coupled with the 3D-FVM radiative transfer method. 
In the equator-on view, the line profiles calculated using the MHD magnetosphere have more high-velocity absorption blueward of line centre, which agrees with our expectation from a deformed dipole. We reiterate that as the magnetosphere becomes larger, the field lines will follow a dipolar topology; thus, for larger \Alf radii, these discrepancies are expected to be minimal.


\subsection{Line Strength}
\label{sec:line_strength}

\begin{figure*}
\centering
\subfigure[$\ra = 2.7~\rs$]{
\includegraphics[width=0.48\textwidth]{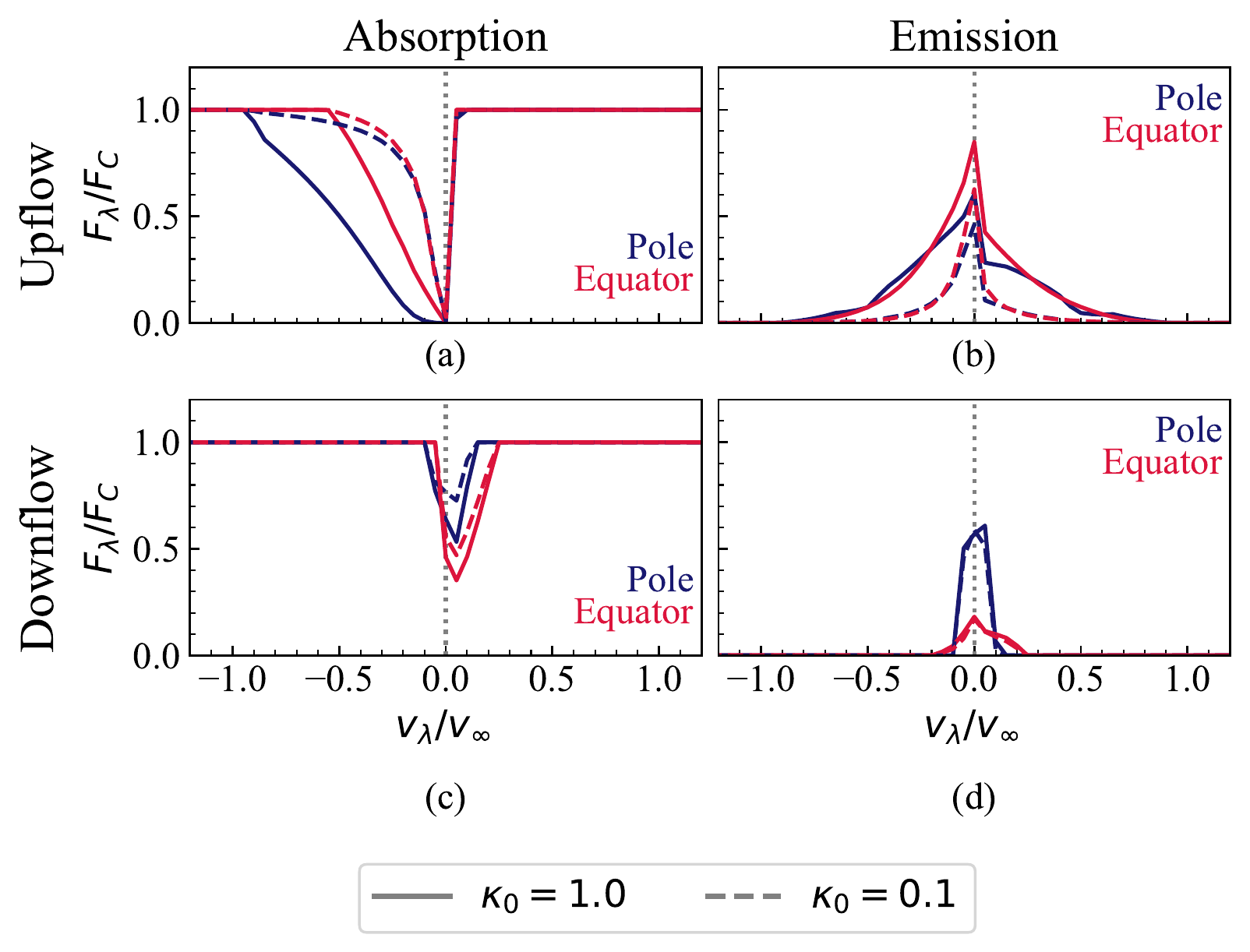}
\label{fig:updown_kappas_2p7}
}
\hfill
\subfigure[$\ra = 10.0~\rs$]{
\includegraphics[width=0.48\textwidth]{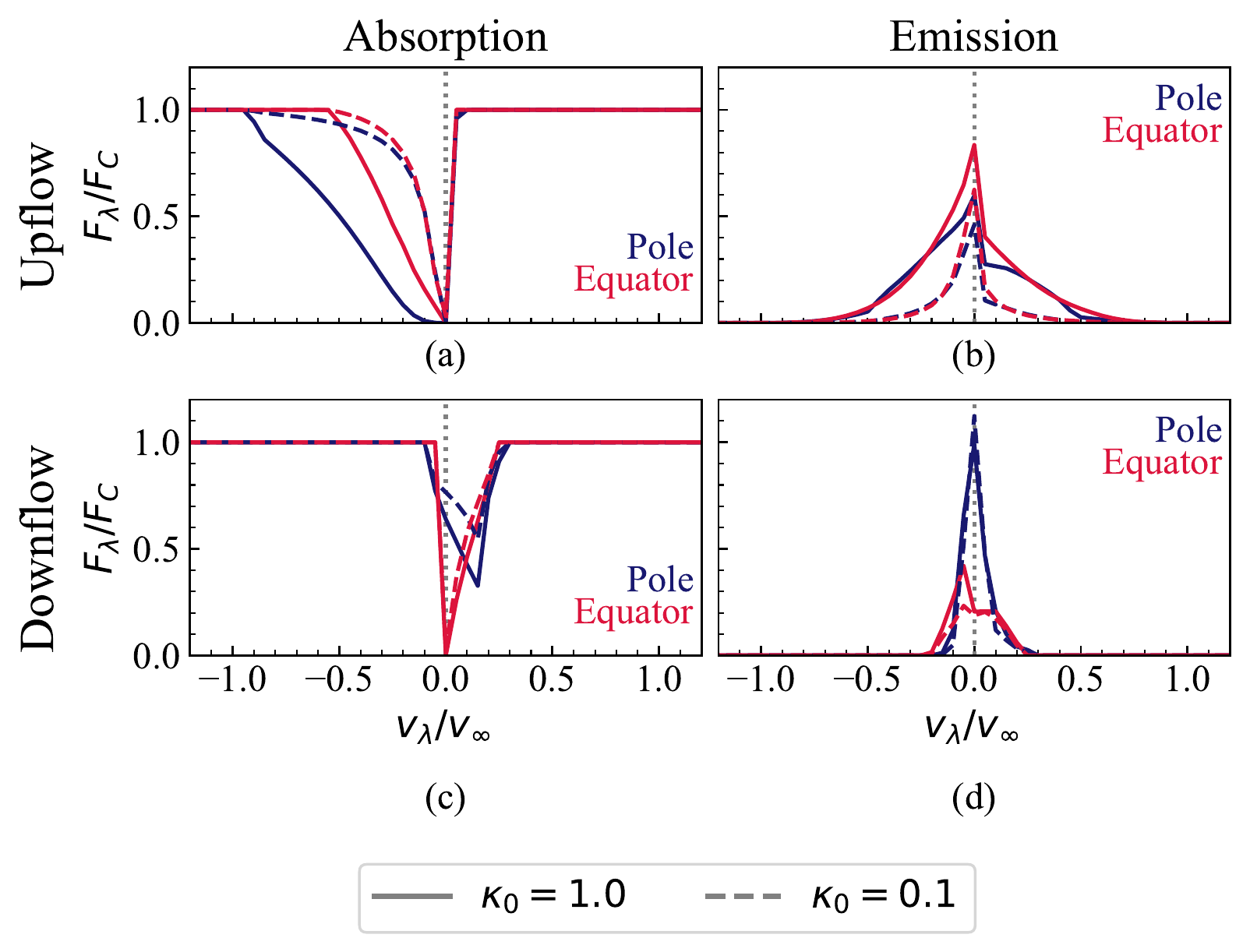}
\label{fig:updown_kappas_10}
}
\caption{Synthetic UV absorption and emission profiles calculated at pole-on (blue lines) and equator-on (red lines) viewing angles, with \Alf radii of $\ra=2.7~\rs$ (left) and $\ra=10.0~\rs$ (right), cooling parameter $\chii=1.0$, and line strengths $\ko = 0.1$ (dashed lines) and $\ko = 1.0$ (solid lines). The upflow (top row) and downflow (bottom row) components show absorption (left) and emission (right) profiles originating exclusively from the upflowing (downflowing) wind. The weak line generally exhibits less absorption and emission for each component. The individual features of the line profile components are discussed more in the text.}
\label{fig:updown_kappas}
\end{figure*}

The shape of the line profile is also affected by the line strength parameter $\ko$ (see Equation \ref{eq:kappa_0}). Following \citet{Marcolino2013}, we compare the synthetic line profiles calculated with $\ko=1.0$ from the previous section with line profiles calculated with $\ko=0.1$. A smaller $\ko$ could represent e.g. a magnetosphere with a lower wind feeding rate, a spectral line with a lower oscillator strength, or an ion with a lower abundance. More specifically, an assessment of which lines are strong or weak is highly dependent on the star and the ionization species under consideration \citep[e.g.][]{Marcolino2012,Marcolino2013}.

\begin{figure*}
\centering
\subfigure[$\ra = 2.7~\rs$]{
\centering
\includegraphics[width=0.48\textwidth]{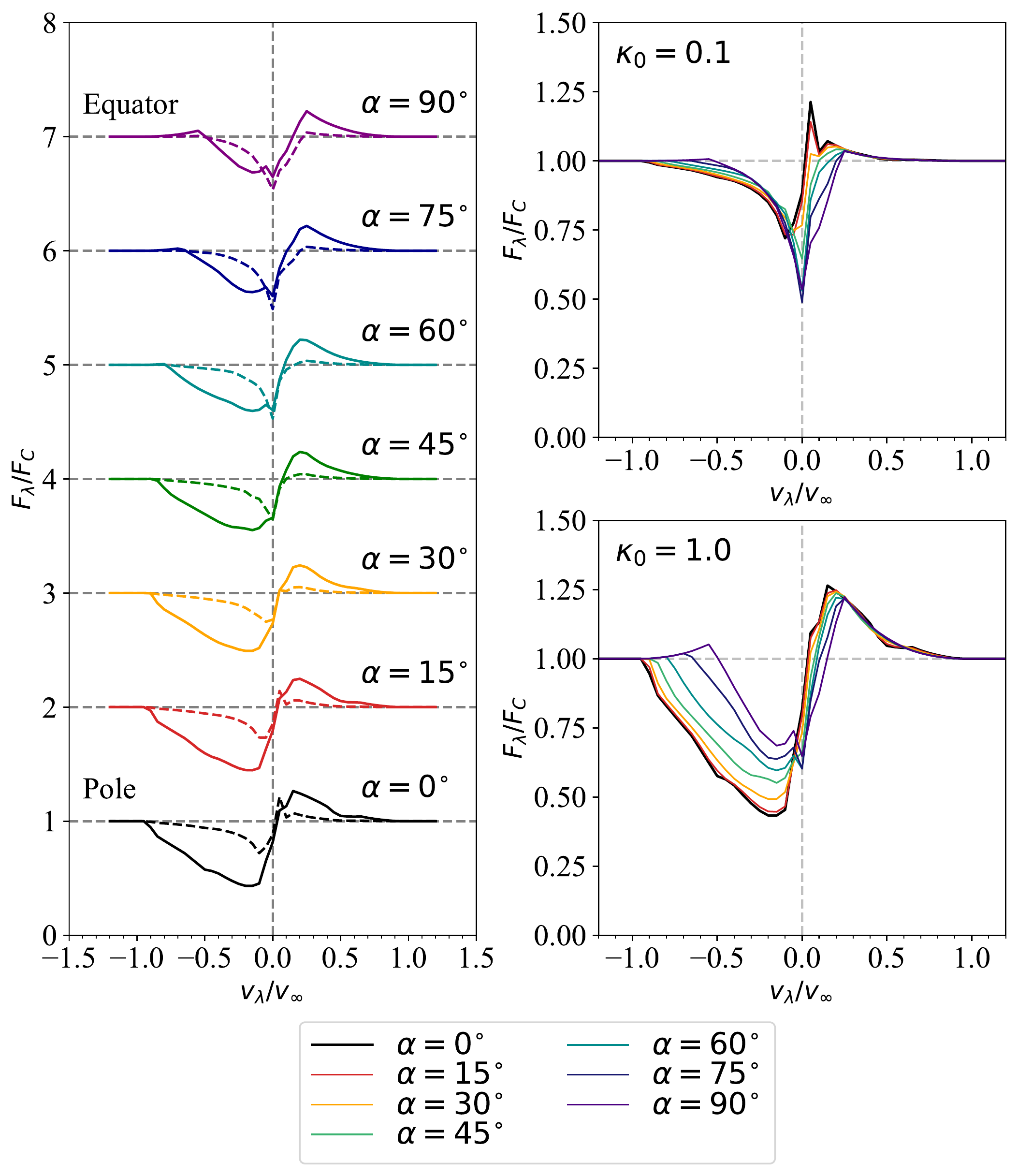}
\label{fig:kappas_2p7}
}
\subfigure[$\ra = 10.0~\rs$]{
\centering
\includegraphics[width=0.48\textwidth]{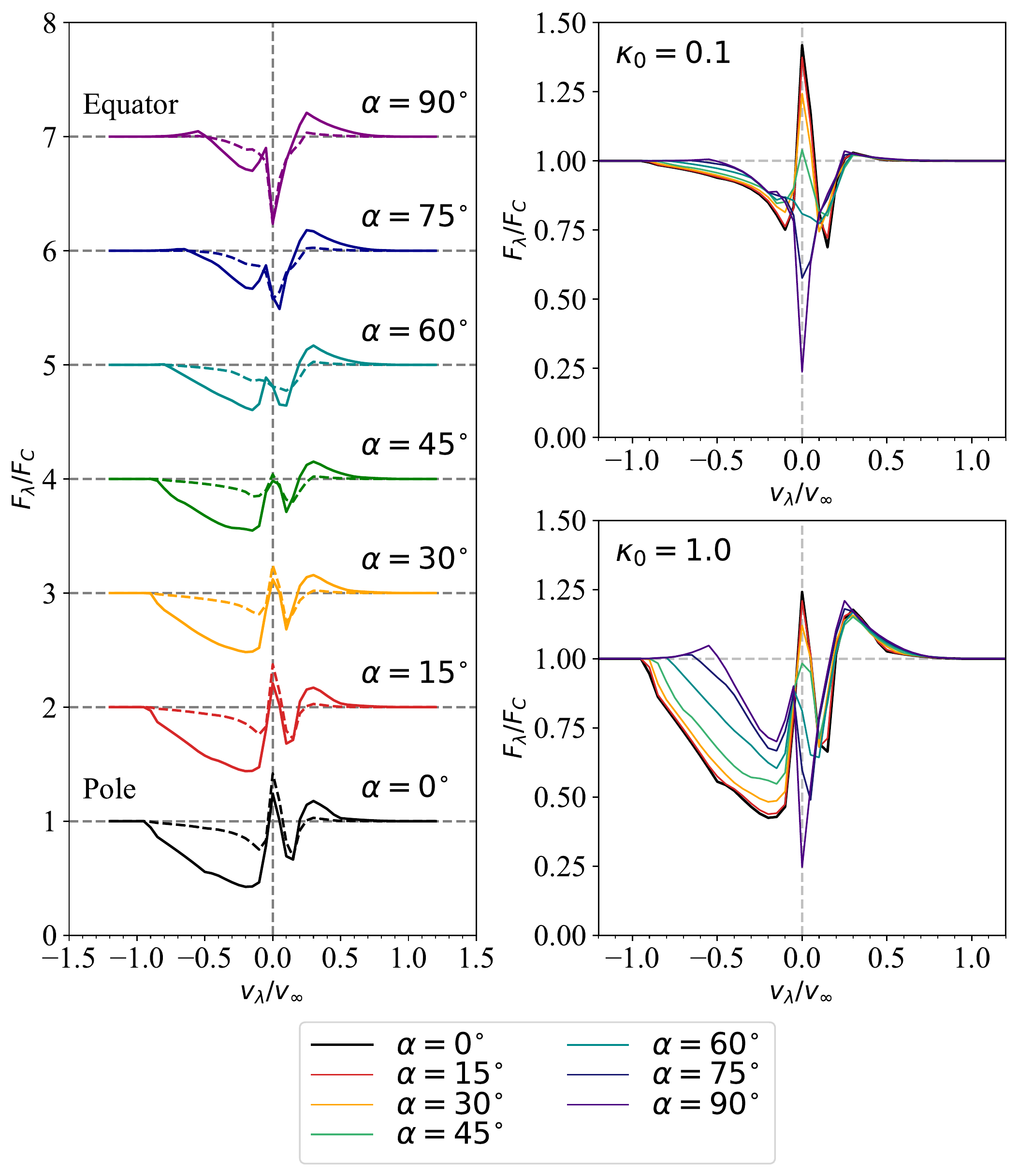}
\label{fig:kappas_10}
}
\caption{Synthetic UV line profiles computed with characteristics similar to those of magnetic O-type stars with (a) $\ra = 2.7~\rs$ and (b) $\ra = 10.0~\rs$. For each \Alf radius, the left panel shows the progression from a pole-on view ($\alpha=0^{\circ}$) to an equator-on view ($\alpha=90^{\circ}$), comparing the profiles for a line strength parameter of $\ko=1.0$ (solid curves) and a line strength parameter of $\ko=0.1$ (dashed curves). The two panels on the right-hand side directly compare the viewing angle variation for the two line strength parameters.}
\label{fig:kappa}
\end{figure*}

Figures \ref{fig:updown_kappas_2p7} ($\ra = 2.7~\rs$) and \ref{fig:updown_kappas_10} ($\ra = 10.0~\rs$) show the absorption (left) and emission (right) components of the absorption profiles caused by the upflow (top) and downflow (bottom) material, for $\ko=0.1$ (dashed lines) and $\ko=1.0$ (solid lines), at pole-on (blue lines) and equator-on (red lines) viewing angles. 

At all \Alf radii and all viewing angles, the absorption part of the $\ko=0.1$ line profiles due to the upflow material (panels a) lacks the absorption at high blue velocity that is present for $\ko=1.0$. The emission profiles due to the upflow material (panels b) similarly have less emission at both high blue and high red velocities compared to the $\ko=1.0$ case. We reiterate that the absorption and emission profiles due to the upflow material does not change with the size of the magnetosphere (see section \ref{sec:views}). Furthermore for $\ko=0.1$, the change with viewing angle is minimal. The weak line is therefore ineffective at probing the high velocity, low density upflow material far from the stellar surface.

For all profiles, the absorption due to the downflow material (panels c) is weaker than the corresponding profiles with $\ko = 1.0$, except for the equator-on view at $\ra = 10.0~\rs$. Here the downflow wind material has a large column density (see density contours in Figure \ref{fig:veq_10}), and therefore has a large optical depth even at low values of $\ko$. Finally, the emission due to the downflow material in all cases does not change significantly with line strength because the optical depth is already large; significant changes in the line profile of the weak line parameter compared to that of the corresponding strong line parameter are therefore mainly due to the upflow wind.

However, for a weak line, the variation of the line profile {\em with viewing angle} is due to the downflow wind. 
As mentioned above, the emission profile due to the upflow does not change significantly between pole-on and equator-on viewing angles. This stands in contrast to the line profile with $\ko = 1.0$, where both the upflow and the downflow wind contribute to the variation of the line profile with changing $\alpha$. The left panels of Figures \ref{fig:kappas_2p7} and \ref{fig:kappas_10} show the variation of the total line profile for the same $\alpha$ as in Figure~\ref{fig:phases}, for $\ko=0.1$ (dashed lines) and $\ko=1.0$ (solid lines), in a star with $\ra = 2.7~\rs$ and $\ra = 10.0~\rs$, respectively. For both \Alf radii, as the star transitions to larger ($\alpha \geq 45^{\circ}$) viewing angles, the emission peak of the weak line is barely visible above the continuum.

\begin{figure}
\centering
\includegraphics[width=0.45\textwidth]{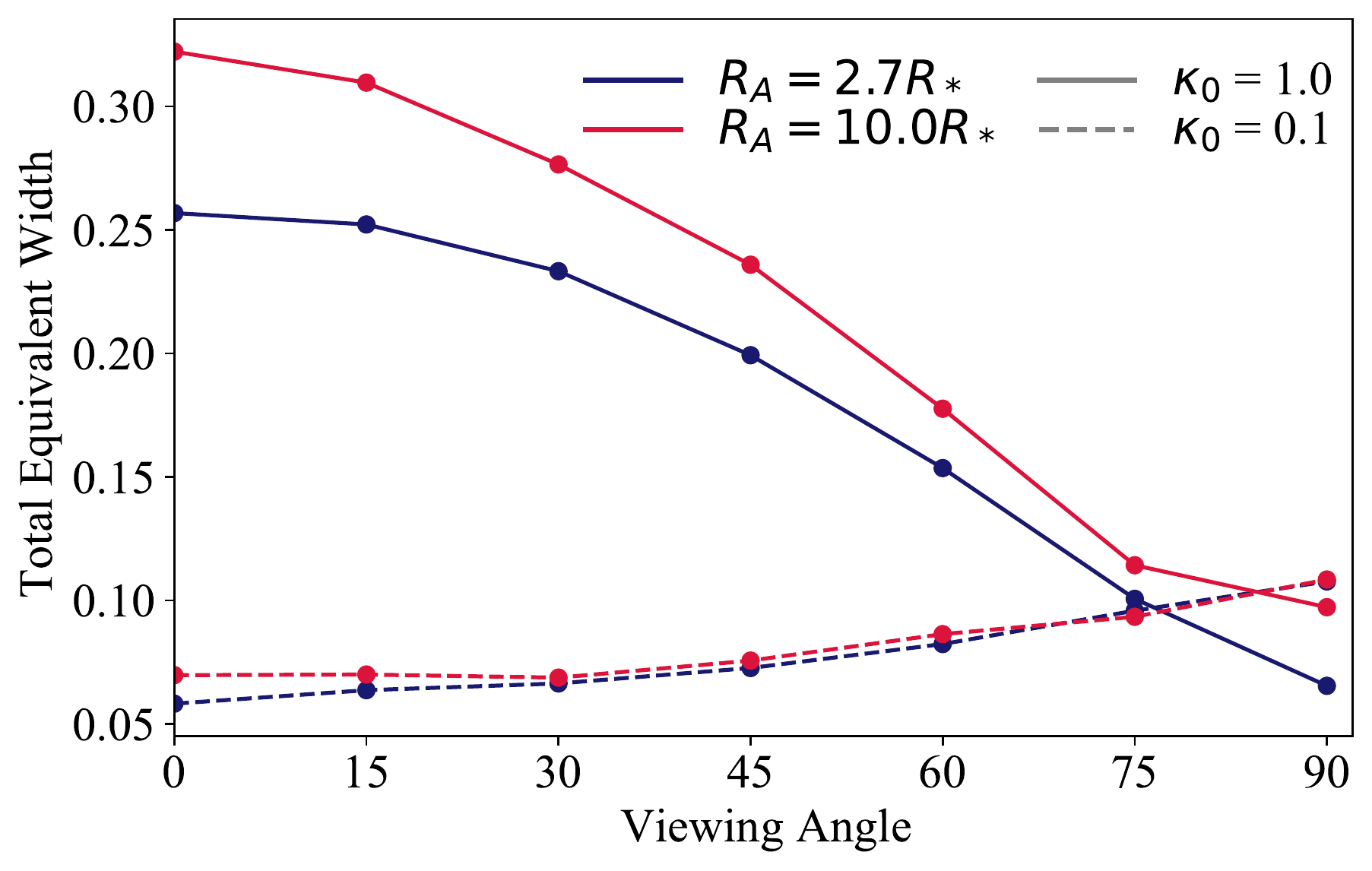}
\caption{Total equivalent widths, as a function of viewing angle, of the line profiles formed for stellar magnetospheres with \Alf radii of  $\ra = 2.7~\rs$ (blue lines) and $\ra = 10.0~\rs$ (red lines), for line strength parameters of $\ko = 0.1$ (dashed lines) and $\ko = 1.0$ (solid lines).}
\label{fig:eqw}
\end{figure}

The right panel of Figures \ref{fig:kappas_2p7} and \ref{fig:kappas_10} reproduces the profiles in the left panel, but with the variation in viewing angle overplotted for the line profile with $\ko = 0.1$ (top panel) and the profile with $\ko = 1.0$ (bottom panel) for $\ra = 2.7~\rs$ and $\ra = 10.0~\rs$, respectively. Unsurprisingly, as shown in Figure \ref{fig:eqw}, we find that the equivalent width of the full line profile (integrated between -1 and 1) of the $\ko = 0.1$ line profiles is significantly less than that of the $\ko = 1.0$ profiles for both $\ra = 2.7~\rs$ and $\ra = 10.0~\rs$ for most viewing angles. However, for $\alpha>75^\circ$, the total equivalent width of the $\ko = 0.1$ profile (dashed lines) becomes larger (more absorption) than that of the $\ko = 1.0$ profile (solid lines). This is because even though the $\ko=1.0$ line profile has more total absorption than the $\ko=0.1$ profile, it also has more emission. 

The equivalent width of the line profile with $\ko = 0.1$ increases as the viewing angle approaches the equator, whereas the equivalent width of the $\ko=1.0$ line profile decreases with viewing angle. This result agrees with that of \citet{Marcolino2013}, who produced synthetic line profiles with $\ko = 0.1$ and $\ko = 1.0$ line parameters at pole-on and equator-on views, using an MHD simulation \citep{udDoula2002,Sundqvist2012} for a star with parameters similar to that of our $\ra = 2.7~\rs$ model. Within the limitation of the ADM formalism, we show that this behavior is the same for larger magnetospheres.

Assuming the geometry of the magnetic field is known, the variation of the line profile with viewing angle therefore puts further constraints on the line strength parameter, breaking potential degeneracy with other ADM parameters when performing line fitting. The line strength parameter depends on known atomic parameters, uncertain ion abundances, and a 
wind-feeding rate which we seek to constrain. However, the degeneracy between the latter two can be lifted, leveraging an ensemble of wind-sensitive lines within a single observation (as the relative ion abundances can be estimated in a consistent manner).


\begin{figure}
\centering
\includegraphics[width=0.47\textwidth]{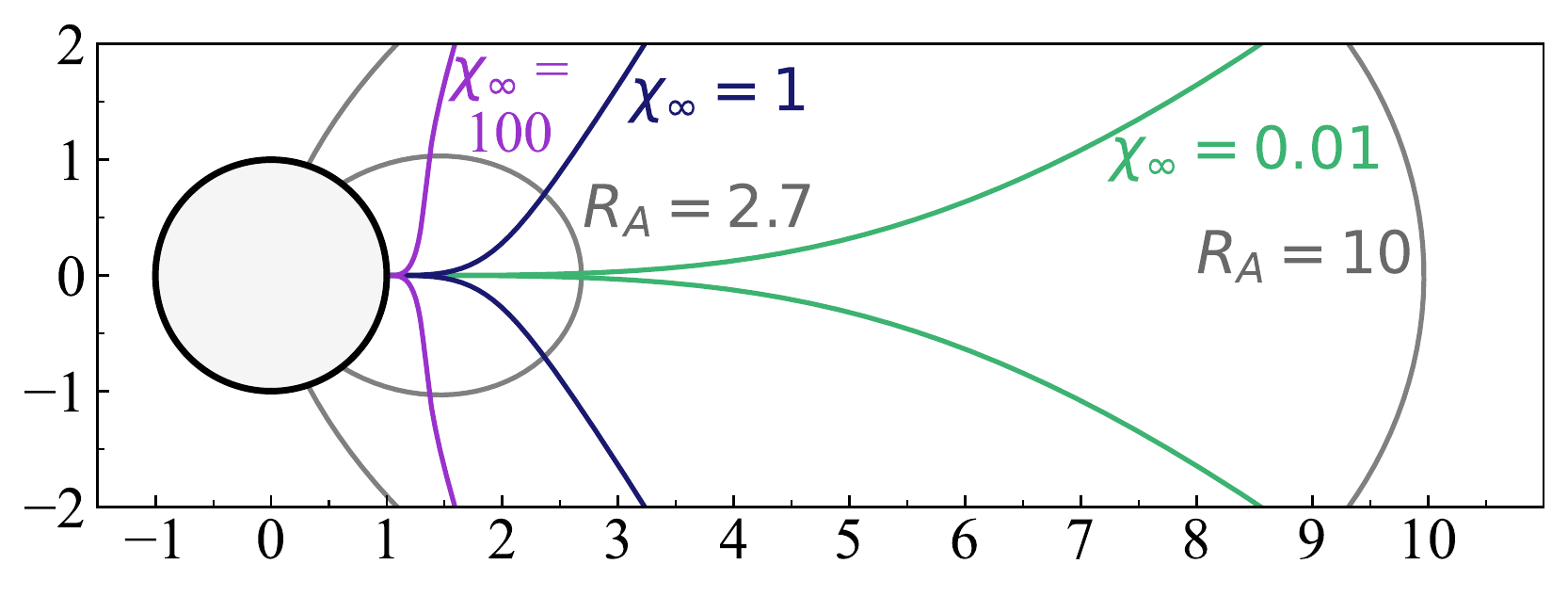}
\caption{A schematic illustrating the effect of a conservative ($\chii = 0.01$), moderate ($\chii = 1.0$), and large ($\chii = 100.0$) cooling parameter, in the case of a small ($\ra = 2.7~\rs$) and a large ($\ra = 10.0~\rs$) \Alf radius (as labelled in the figure). For clarity, x and y-axes are measured in units of $\rs$. The potential impact of the cooling parameter is considerably more significant for stars with large magnetospheres.} 
\label{fig:duckplot}
\end{figure}

\begin{figure*}
\centering
\subfigure[$\ra = 2.7~\rs$]{
\centering
\includegraphics[width=0.48\textwidth]{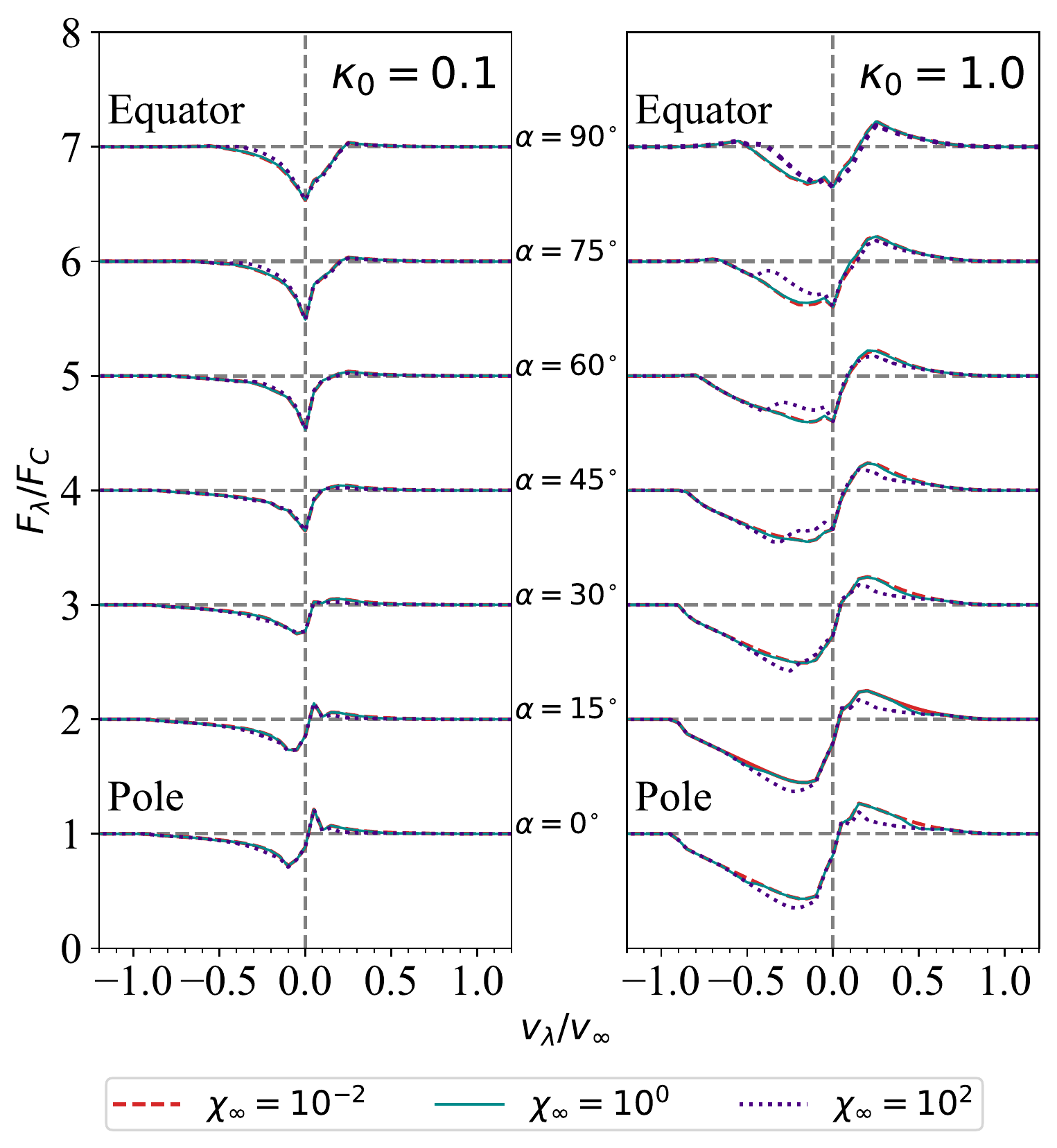}
\label{fig:chii_2p7}
}
\subfigure[$\ra = 10.0~\rs$]{
\centering
\includegraphics[width=0.48\textwidth]{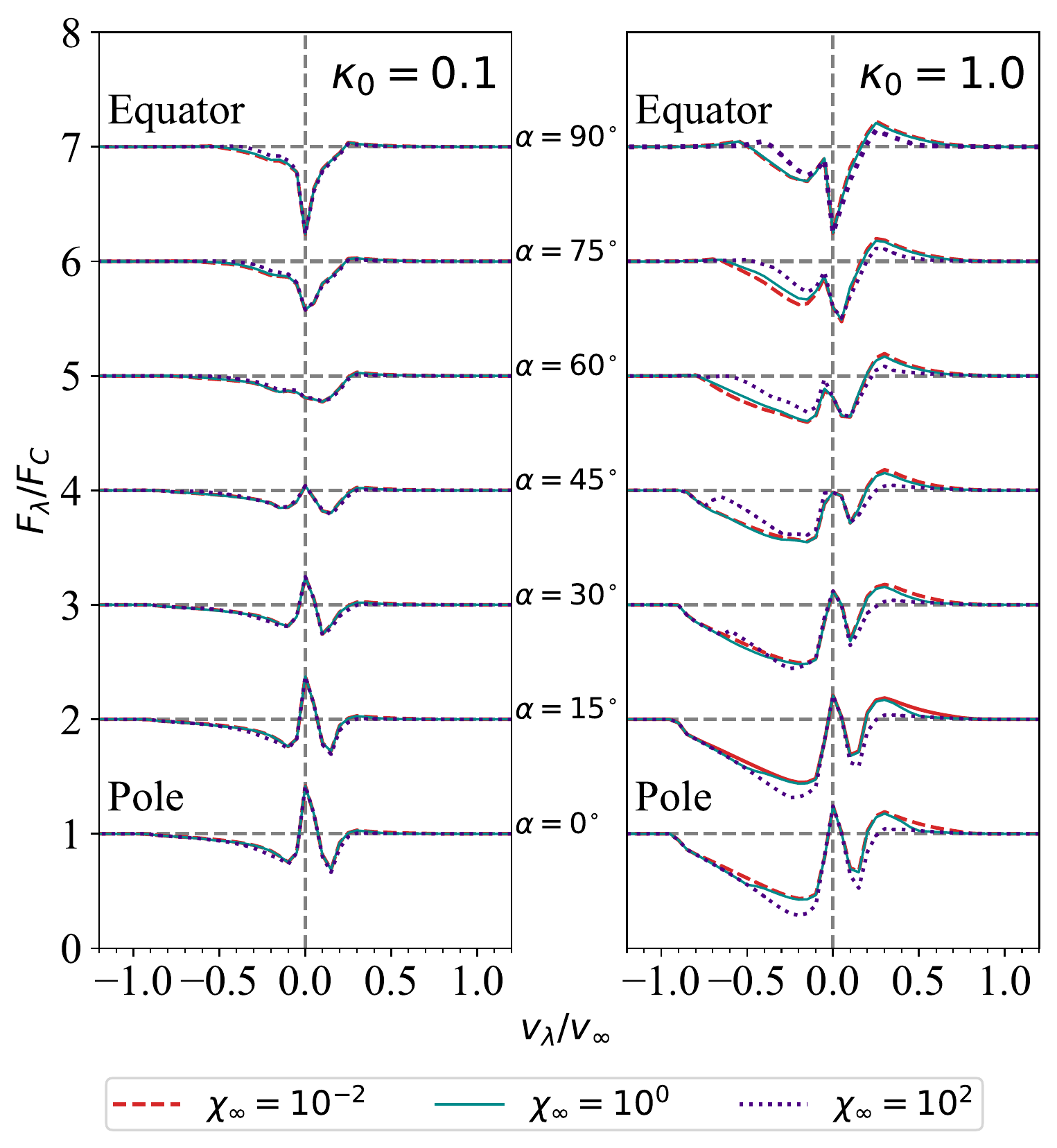}
\label{fig:chii_10}
}
\caption{Synthetic UV line profiles computed with characteristics similar to those of magnetic O-type stars with \Alf radii $\ra = 2.7~\rs$ (a) and $\ra = 10.0~\rs$ (b). For each \Alf radius, the left panel shows the progression from a pole-on view ($\alpha=0^{\circ}$) to an equator-on view ($\alpha=90^{\circ}$) for a line strength parameter of $\ko=0.1$, comparing three different cooling parameters ($\chii = 0.01, \chii = 1.0, \chii = 100.0$) at each viewing angle. The right panel shows the same figure, but computed for a line strength parameter of $\ko=1.0$.}
\label{fig:chiinf}
\end{figure*}

\subsection{Cooling Parameter and Smoothing Length}
\label{sec:cool_param}

We also consider here the effect of the cooling parameter ($\chii$, see Equation \ref{eq:chii}) on the line profiles. As described above in Section \ref{sec:ADM}, in the ADM formalism, the hot post-shock gas in the upflow is essentially transparent in the UV, and the extent of this region is parametrized by the cooling parameter $\chii$. Figure \ref{fig:duckplot} shows the shock boundary location corresponding to a small ($\chii = 0.01$), moderate ($\chii = 1.0$), and large ($\chii = 100.0$) cooling parameter.

Figure \ref{fig:chiinf} shows synthetic line profiles calculated at $\ra = 2.7~\rs$ (a) and $\ra = 10.0~\rs$ (b), with line strength parameters $\ko=0.1$ (left panel of each subfigure) and $\ko=1.0$ (right panel of each subfigure), for three different cooling parameters (overplotted), at viewing angles progressing from pole-on to equator-on. The shape of the line profile changes slightly when $\chii$ becomes very large, but only for the line profiles with $\ko=1.0$. This change is more pronounced for magnetospheres with larger \Alf radii, but is probably still barely perceptible at the typical signal-to-noise ratios of hot star UV spectroscopy.

As can be seen from the density contours in Figure \ref{fig:vplot}, the hot post-shock gas region primarily removes upflow material that would have had low density. Because of this, the cooling parameter is therefore limited in its usefulness to diagnose wind properties such as, e.g. the mass-loss rate. Thus overall, although the cooling parameter ranges by four dex in our models, \textit{there is no significant change to the line profiles} at either a moderate or extended \Alf radius at either line strength.


\begin{figure}
\centering
\includegraphics[width=0.45\textwidth]{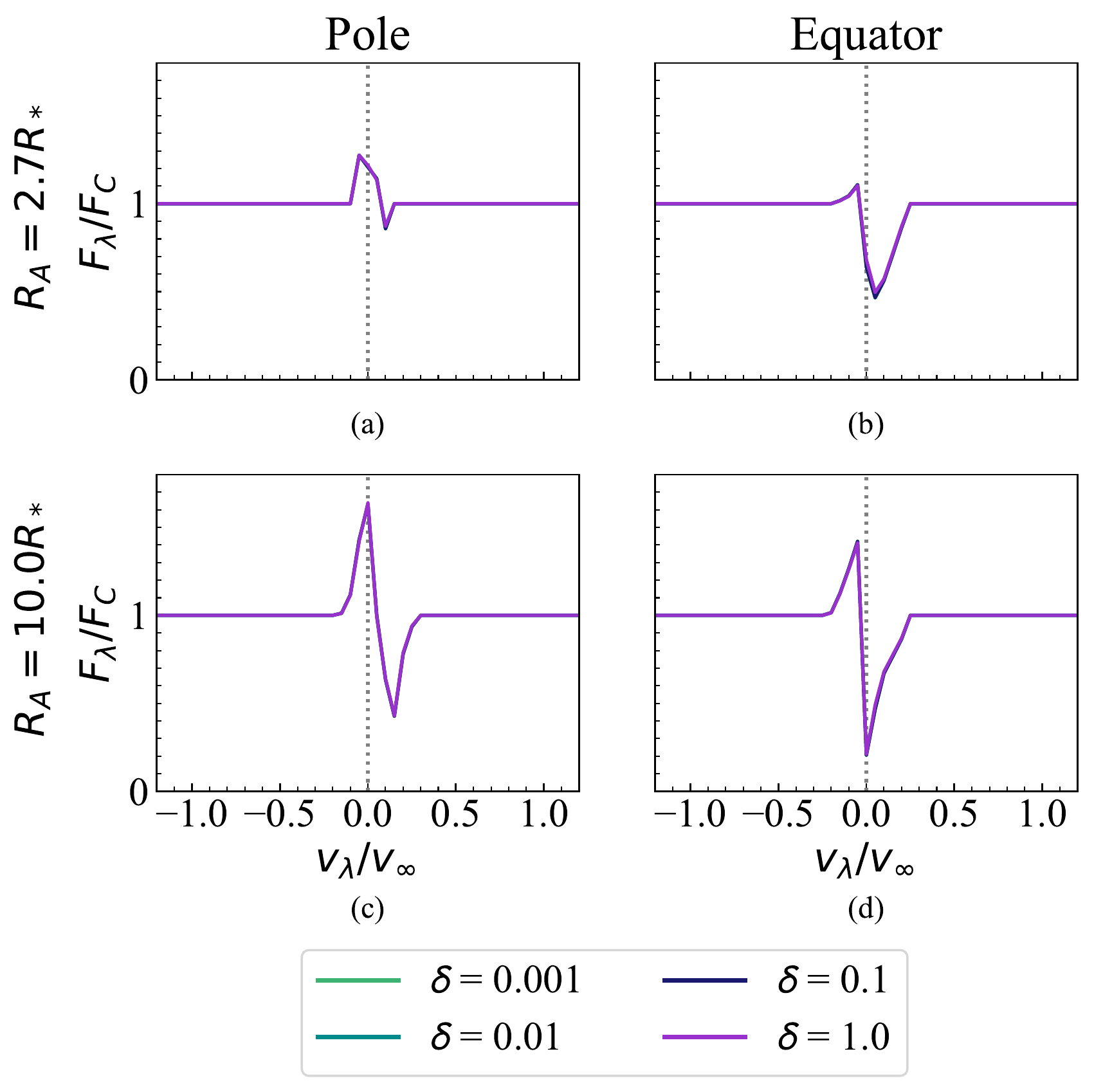}
\caption{Contribution of the downflow component to the synthetic UV line profiles, calculated at $\ra = 2.7~\rs$ (top) and $\ra = 10.0~\rs$ (bottom), with line strength parameter $\ko=1.0$ and cooling parameter $\chii = 1.0$, at both pole-on (left panel of each subfigure) and equator-on (right panel of each subfigure) viewing angles, for smoothing factors $\delta/\rs = [0.001,0.01,0.1,1.0]$.}
\label{fig:deltas}
\end{figure}

The smoothing length ($\delta$; \citetalias[Equation 24]{Owocki2016}) is a spatial smoothing factor scaled to the stellar radius in the downflow region. Since $\delta$ exclusively affects downflow material, Figure \ref{fig:deltas} shows only the downflow contribution to the line profile, for $\ra = 2.7~\rs$ (top) and $\ra = 10.0~\rs$ (bottom), with line strength parameter $\ko=1.0$ and cooling parameter $\chii = 1.0$, at both pole-on (left panel of each subfigure) and equator-on (right panel of each subfigure) viewing angles, for four smoothing lengths $\delta/\rs = [0.001,0.01,0.1,1.0]$. Even with this large variation in the size of the smoothing length, there is {\em no noticeable change} in the shape of the line profiles. This indicates that $\delta$ would have to be quite large (at least on the order of a few stellar radii) before it measurably impacts the total line profile.

\subsection{Turbulent Velocity}
\label{sec:microturb}

\begin{figure}
\centering
\includegraphics[width=0.47\textwidth]{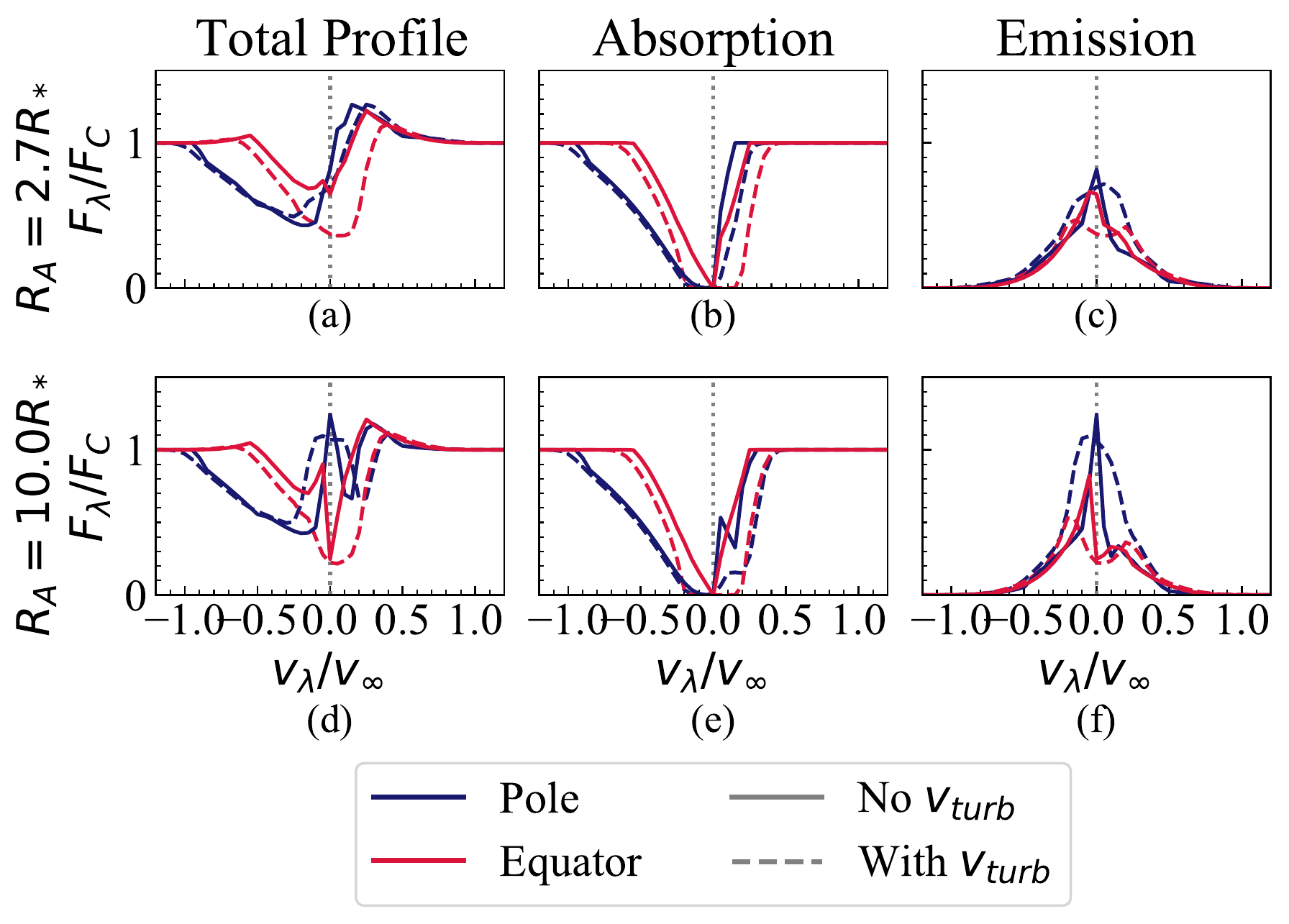}
\caption{Synthetic UV line profiles calculated at $\ra = 2.7~\rs$ (top) and $\ra = 10.0~\rs$ (bottom), with line strength parameter $\ko=1.0$ and cooling parameter $\chii = 1.0$, at both pole-on (blue lines) and equator-on (red lines) viewing angles. Solid lines indicate profiles calculated with thermal broadening only (these are the same as the profiles from Section \ref{sec:views}). Dashed lines indicate profiles calculated with $v_{\text{turb}} = 0.1 \vinf$. Each row contains three panels showing the full line profile (left), as well as the individual absorption (middle) and emission (right) components.} 
\label{fig:large_vth}
\end{figure}

\citet{Sundqvist2012} used 2D MHD simulations of dynamical magnetospheres to model the equivalent width variations of the magnetic star HD~191612. The authors found that the addition of a turbulent velocity term to the profile function on the order of 100~km~s$^{-1}$ was required to reproduce the observed shape of the H$\alpha$ line profile. \citetalias{Hennicker2018} also employed turbulent velocity parameters of 100~km~s$^{-1}$ (for calculating the source function) and 50~km~s$^{-1}$ (for calculating the line profiles) in their models calculated using an MHD magnetosphere. 
While we do not generally include turbulent broadening in our synthetic line profiles, we include here a brief discussion of its effect on the line profiles presented in Section~\ref{sec:views}.

Figure~\ref{fig:large_vth} shows the synthetic UV line profiles calculated at $\ra = 2.7~\rs$ (top) and $\ra = 10.0~\rs$ (bottom), with line strength parameter $\ko=1.0$ and cooling parameter $\chii = 1.0$, at both pole-on (blue lines) and equator-on (red lines) viewing angles. The line profiles from Section~\ref{sec:views} are shown in solid lines; these do not include a turbulent velocity. We then reproduce those profiles, increasing the thermal velocity term in the profile function (Equation~\ref{eq:prof_func}) to $\vth = 0.1 \vinf$, mimicking the addition of a turbulent velocity component that adds to the thermal broadening (dashed lines in Figure~\ref{fig:large_vth}). We note that this does not affect the source function calculation in the optically thin regime. Each row of the figure contains three panels showing the full line profile (left), the absorption profile (middle), and the emission profile (right) for each model, in order to illustrate the effect of turbulent broadening on each component of the line profile. 

The addition of turbulent velocity broadens and strengthens the line profile because of the wider resonance zones. However, such changes do not depend on the size of the magnetosphere. The variations in the synthetic line profiles for magnetospheres with $\ra =2.7~\rs$ and $\ra =10.0~\rs$ are qualitatively the same.

In principle, a turbulent velocity that changes with position in the atmosphere could be added to our models, but the physical basis for this variation with position is still missing. New 3D MHD simulations of obliquely rotating magnetospheres \citep{udDoula_MOBSTER1}, and new 2D MHD simulations of magnetized stars with Line-Deshadowing Instabilities included \citep{Driessen_MOBSTER1} will provide important constraints on future modeling efforts.

\section{Conclusions}
\label{sec:conclusions}

In this paper, we present a systematic investigation of the formation of UV resonance lines in the magnetospheres of massive stars. We produce synthetic spectra by pairing the ADM formalism with a radiative transfer technique that leverages the optically thin source function to perform exact integration. We examine synthetic spectra for seven viewing angles between a pole-on and an equator-on view, for several combinations of the values of the \Alf radius, line strength and cooling parameter.

Overall, we confirm that both the upflow and downflow components of the ADM contribute to the unusual shape of the line profiles in dynamical magnetospheres. In particular, magnetic massive stars uniquely exhibit redshifted absorption. This is mainly due to the downflow wind, although there is also a small but non-negligible contribution from the upflow wind. This phenomenon is not observed in stars with spherically symmetric winds, therefore redshifted absorption is strongly indicative of the presence of a magnetic field. Additionally, we show that the amount of redshifted absorption due to the downflow increases as the viewing angle shifts from pole-on to equator-on views.

Unlike the broad and smooth emission profile resulting from a spherically symmetric wind, the synthetic emission profiles for a magnetic wind are broad at higher velocities with a narrow peak near line centre that is due to the downflow wind component. This asymmetry in the emission profile cannot be explained by the occultation effects of spherically symmetric models. 
Although the variation of the emission profile with viewing angle is small (when compared to that of the absorption), the low-velocity emission peak contributes to the atypical shape of the line profile, especially in magnetospheres with strong magnetic fields. 

The line strength parameter, which is proportional to the mass-loss rate of the wind and to the atomic parameters of a specific line, also impacts the shape of the line profile. We confirm that the line profile with a line strength parameter of $\ko=0.1$ is ineffective at probing the high velocity, low density upflow material far from the stellar surface, therefore the absorption profiles of the weak line lack the characteristic extended blue edge seen in those of the $\ko=1.0$ profile. Additionally, we find that significant differences in the line profile of the weak line parameter compared to that of the strong line parameter are mainly due to the upflow wind component. For weak lines, however, the variation of the line profile with viewing angle is largely due to the downflow wind. In contrast, both the upflow and downflow wind components contribute to the variation of stronger lines (e.g. $\ko=1.0$) with viewing angle.

We show that the cooling parameter, which defines the location of the hot post-shock material, has a negligible effect on the line profiles regardless of the strength of the line, the value of the \Alf radius, or the viewing angle between the observer's direction and the magnetic field axis. Indeed, in the case of a line strength parameter of $\ko=1.0$ and a large \Alf radius, the line profiles are only mildly affected by cooling parameter. Similarly, the smoothing length was also shown to have no significant impact on the line profiles examined.

Finally, we find that a large (high) velocity dispersion, estimated by introducing a turbulent velocity term, yields broader and stronger line profiles when compared to a smaller (lower) velocity dispersion. These differences are independent of the size of the magnetosphere. 

Our results show that synthetic line profiles generated using the ADM formalism coupled with radiative transfer techniques may provide a useful new approach for assessing the observed behavior of the wind line profiles of slowly rotating magnetic massive stars, {\em without} the computational cost of using MHD simulations or computationally challenging self-consistent radiative transfer methods. We expect our synthetic line profiles will aid in the interpretation of observational data and in providing direct constraints on the properties of massive star magnetospheres (e.g. the ADM-based models of photometric variability by \citealt{Munoz2020}). The parameter study presented here addresses singlet lines, which can be compared to singlet lines or well-spaced doublets in observed spectra. Extensions of this method that will enable the modeling of doublet lines will be presented in a future study.

Finally, our models can also determine specific spectral features in the UV that might be unique to magnetic stars. This will prove particularly useful in light of large observational surveys such as the UV Legacy Library of Young Stars as Essential Standards (ULLYSES) project\footnote{\url{https://ullyses.stsci.edu/}}, which will produce UV spectral libraries of O- and B-type stars in the (Large and Small) Magellanic Clouds \citep{ullyses2020}.

\section*{Acknowledgements}

CE gratefully acknowledges support for this work provided by NASA through grant number HST-AR-15794.001-A from the Space Telescope Science Institute, which is operated by AURA, Inc., under NASA contract NAS 5-26555. CE also gratefully acknowledges graduate assistant salary support from the Bartol Research Institute in the Department of Physics and Astronomy at the University of Delaware.

CE and VP gratefully acknowledge support for this work provided by NASA through grant numbers HST-GO-15066, HST-GO-13734, and HST-GO-13629 from the Space Telescope Science Institute, which is operated by AURA, Inc., under NASA contract NAS 5-26555.

VP gratefully acknowledges support from the University of Delaware Research Foundation.

ADU gratefully acknowledges support from the Natural Sciences and Engineering Research Council of Canada (NSERC). This work is supported by NASA under award number 80GSFC17M0002.

LH and JOS gratefully acknowledge support from the Odysseus program of the Belgian Research Foundation Flanders (FWO) under grant G0H9218N.

YN acknowledges support from the Fonds National de la Recherche Scientifique (Belgium), the European Space Agency (ESA) and the Belgian Federal Science Policy Office (BELSPO) in the framework of the PRODEX Programme linked to XMM-Newton.

AuD acknowledges support by NASA through Chandra Award number TM1-22001B issued by the Chandra X-ray Observatory Center, which is operated by the Smithsonian Astrophysical Observatory for and behalf of NASA under contract NAS8-03060.

The authors wish to thank Dr. Stan Owocki for his helpful comments during the early stages of this project. The authors would also like to thank the anonymous referee for their thoughtful review of the manuscript.

\section*{Data Availability Statement}

The UV-ADM code, as well as the grid of models produced for and used in this work, are available from the authors upon request. 


\bibliographystyle{mnras}
\bibliography{database_Erba} 


\appendix

\bsp	
\label{lastpage}
\end{document}